\documentclass[usenatbib]{mn2e} % Normal MNRAS format
\usepackage{longtable} % Allows table to span multiple pages
\usepackage{amssymb, textcomp, verbatim, graphicx, times, multirow} % hyperref, 

\def\aj{AJ}% % Astronomical Journal
\def\araa{ARA\&A}% % Annual Review of Astron and Astrophys
\def\apj{ApJ}% % Astrophysical Journal
\def\apjs{ApJS}% % Astrophysical Journal, Supplement
\def\aa{A\&A}% % Astronomy and Astrophysics
\def\aas{A\&AS}% % Astronomy and Astrophysics, Supplement
\def\mnras{MNRAS}% % Monthly Notices of the RAS
\def\pasp{PASP}% % Publications of the ASP
\title{Study of X-ray emission from the old open cluster, M67}
\author[Mooley \& Singh]{K.~P.~Mooley $^1$ and K.~P.~Singh$^2$ \\
  $^{1}$ California Institute of Technology, 1200 E. California Blvd., MC 249-17, Pasadena, CA 91125\\
  $^{2}$ Tata Institute of Fundamental Research, Mumbai 400 005, India}

\begin{document}
\maketitle

%--------------------------------------------------------------------------------------------
% ABSTRACT AND KEYWORDS
%--------------------------------------------------------------------------------------------
\begin{abstract}
We present an X-ray analysis of a 4 Gyr old open cluster, M67, using archival {\it XMM-Newton} data. 
The aim of this study was to find new X-ray members of M67,  and to use the updated member list for studying X-ray variability, 
and derive the X-ray luminosity functions (XLFs) of different stellar types and compare them with other star clusters of similar age.
We report the detection of X-ray emission from 25 members of M67, with membership based primarily on their proper motion, of which  
one X-ray source is a new member. Supplementing this study with previous {\it ROSAT} and {\it Chandra} 
studies of M67, and using the most recent proper motion study by \citeauthor{vereshchagin14}, we have compiled a revised list of 
X-ray emitting members of M67 consisting of 43 stars.
Sixteen of these are known RS CVn type binaries having orbital periods $<$ 10 days, and near-circular orbits,
5 are contact binaries with orbital periods $<$ 6 hours, 5 are yellow and blue stragglers, 2 are Algol-type binaries, and one source 
is a cataclysmic variable.
Fourteen members do not have any orbital information and cannot be classified.
Fourteen of the X-ray sources detected do not have any optical counterpart down to a magnitude of $V\simeq22$, and their membership is uncertain.
Finally, we report the X-ray luminosity functions of RS CVn type and other types of stars in M67 and compare them with other 
open clusters of intermediate-to-old age.
% The catalclysmic variable, EU Cnc, previously thought to be a member of M67, is found to be a non-member based on proper motion and other membership tests. 

\end{abstract}

\begin{keywords}
       stars : activity -- 
       binaries : general -- 
       stars : blue stragglers --
       open clusters and associations: individual: M67 -- 
       X-rays : binaries 

\end{keywords}

%--------------------------------------------------------------------------------------------
% INTRODUCTION
%--------------------------------------------------------------------------------------------
\section{INTRODUCTION}\label{sec:intro}

Open clusters are useful for studying coeval and comoving populations of stars within the Galactic disk. 
X-ray studies of stars in open clusters offer an insight into their coronal activity and/or accretion phenomena.
As clusters age, the spin down of stars causes X-ray emission to diminish in general, and thus, in the X-rays, revealing 
active coronae primarily from stars spun-up in binary systems 
\citep[the age-rotation-activity correlation; e.g. ][]{pallavicini89,randich97,gudel04} or from systems undergoing accretion. 
Accordingly, X-rays from old (a few Gyr or older) open clusters are unique probes of magnetically active (RS CVn, BY Dra, W UMa, FK Com, Algol) and 
mass-transfer (CVs, L/HMXBs) binary systems within the Galactic disk \citep{belloni98,verbunt99,berg13a}.

Past studies of old open clusters, NGC 6791 \citep[$\sim$8 Gyr; ][]{berg13}, NGC 188 \citep[$\sim$6 Gyr; ][]{belloni98}, and M67 \citep{belloni93,belloni98,berg04}, 
reveal that RS CVns, CVs, and sub-giant stars dominate the X-ray emission, while peculiar objects such as blue and 
yellow stragglers are rare and thus do not contribute much.
%What NEW  you hope to learn from the X-ray study of M67 ?

M67 is a 4.2$\pm$0.6 Gyr-old open cluster at 850$\pm$30 pc having a small reddening value 
(E$_{B-V}\approx0.04$) \citep{sarajedini1999,yadav08}.
On account of the extensive optical data available for M67, this cluster is well suited for study in X-rays.
Proper-motion studies of M67 to establish cluster membership have been carried out by several groups \citep{sanders77,girard89,zhao93,yadav08,vereshchagin14}.
In this paper we use cluster membership information from \cite{vereshchagin14}, a revised version of the \cite{yadav08} 
catalog reaching down to V$\sim$22 mag, containing 659 members.
Two published X-rays studies of M67 exist.
\cite{belloni93,belloni98} presented {\it ROSAT} PSPC observations covering $\sim$0.5 deg$^2$ of the M67 field, while 
% with the faintest detected source having an X-ray flux $9\times10^{-15}~\mbox{erg cm}^{-2}~\mbox{s}^{-1}$
\cite{berg04} analyzed {\it Chandra} ACIS observations covering $\sim$0.1 deg$^2$ but with a limiting flux of about 40 times lower 
than the {\it ROSAT} observations.

Here, we present the results of two {\it XMM-Newton} observations of M67 having fields of view and limiting flux intermediate 
to that afforded by the  {\it ROSAT} and {\it Chandra} observations.
%The motivation of this study is to find new X-ray members of M67 and compare the X-ray luminosity function of RS CVn type and other
%stellar types in M67 with that in other star clusters of similar age.
% of about $1.6\times10^{-15}$ and $4.8\times10^{-15}~\mbox{erg cm}^{-2}~\mbox{s}^{-1}$ respectively.
The paper is organised as follows.
In Section 2, we discuss the X-ray data and processing. The optical, {\it ROSAT} and {\it Chandra} counterparts of our X-ray sources are given in Section 3. 
The spectral hardness and variability analysis is in Section 4. 
Section 5 gives notes on individual classes of X-ray sources, and we conclude with a discussion in Section 6.

%--------------------------------------------------------------------------------------------
% DATA REDUTION
%--------------------------------------------------------------------------------------------
\section{ARCHIVAL DATA AND DATA PROCESSING}\label{sec:data}

%%%%%%%%%%%%% DATA
\subsection{X-ray Data}\label{sec:data.data}
\begin{figure}
\includegraphics[width=3.4in,viewport=26 10 480 420,clip]{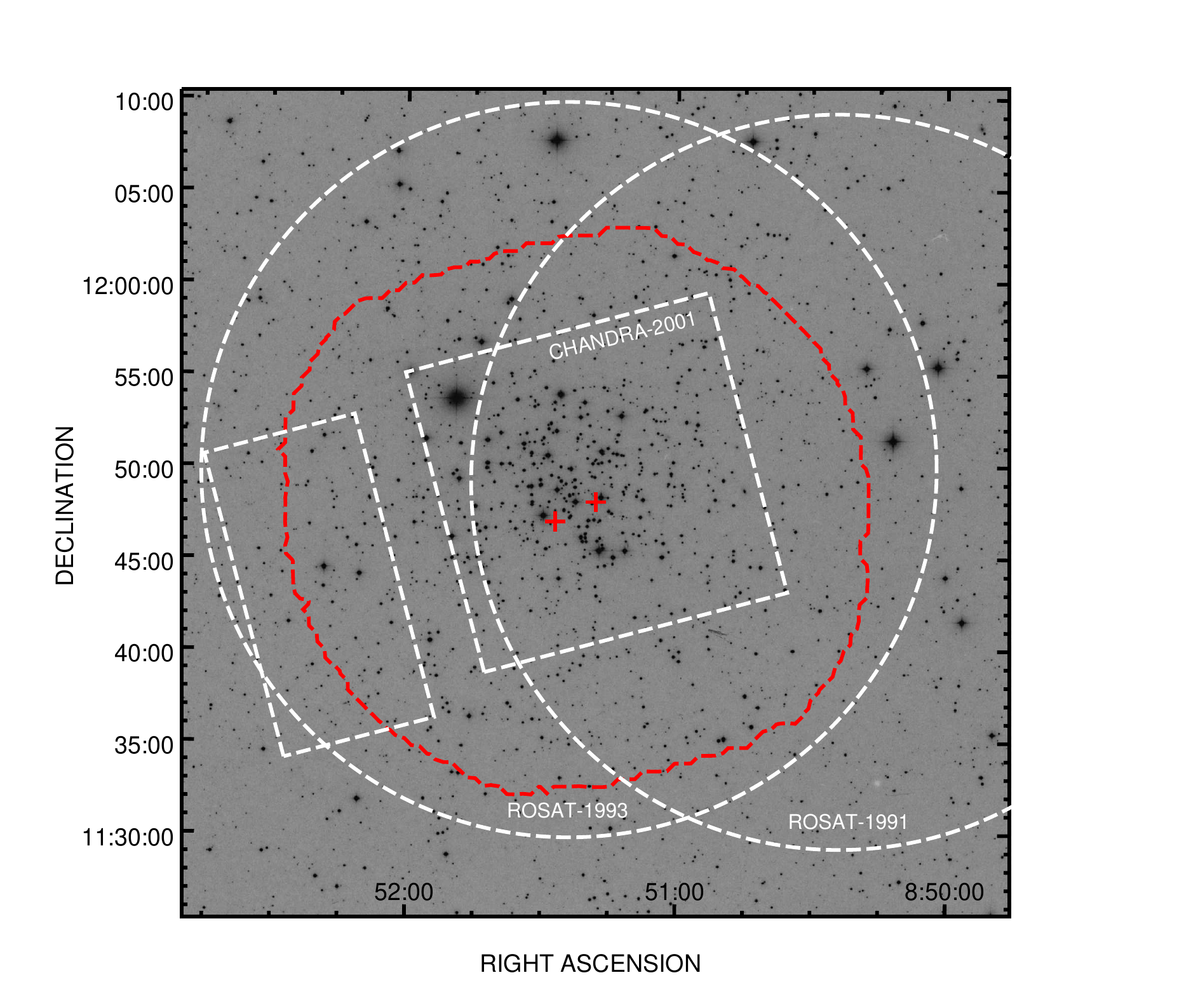}
\caption{A 45\arcmin $\times$ 45\arcmin image of M67 from the Digital Sky Survey.
The footprints of the Chandra ACIS observation \citep[dashed white boxes; ][]{berg04}, ROSAT observations \citep[dashed white circles; ][]{belloni93,belloni98}, 
and the XMM--Newton observations used in this work (red; aimpoints marked by crosses) are overlayed.}
\label{fig:fov}
\end{figure}

We used archival data  from two observations of M67 from the XMM-Newton Science Archive (XSA).
Observation with ID 0109461001 was carried out in 2001 and ID 0212080601 in 2005.
For both the observations, we used the European Photon Imaging Camera \citep[EPIC, consisting of two MOS and one PN CCD arrays; ][]{jansen01,struder01,turner01}. 
The first observation used the thin filter while the second observation used the thick filter. 
Data were acquired in the full frame mode in both the cases.
The observation details are given in Table~\ref{tab:xmmlog}.
The merged {\it XMM-Newton} footprint for the two M67 observations is shown in Figure~\ref{fig:fov}.
The footprints of the previously published {\it ROSAT} and {\it Chandra} observations of M67 are also shown.
Although the 3XMM pipeline products (Watson et al. 2014, in prep) are available, we manually reduced and inspected 
the data to use the latest calibration files and a better control over the filtering, reduction and source selection process.
We downloaded Observation data files (ODFs) from the XSA for further processing.

% ---------------------------------------------------------------
% FIGURE: FOV OF OBSERVATIONS
% ---------------------------------------------------------------
% ---------------------------------------------------------------

% ---------------------------------------------------------------
% TABLE: OBS LOG
% ---------------------------------------------------------------
\begin{table*}
\centering
\caption{XMM-Newton data on M67}
\begin{tabular}{lll} % 3 cols
\hline \hline
Obs ID                          & 0109461001           & 0212080601           \\
Coordinates                     & 08:51:26.99~11:46:~58.0 & 08:51:18.00~11:48:02.6\\
Start time  (UT)                & 20 Nov 2001 23:56:16 & 08 May 2005 18:23:17  \\
Stop time   (UT)                & 21 Nov 2001 02:43:07 & 08 May 2005 22:23:28  \\
Usable time (MOS1,MOS2,PN; ks)  & 9.42,9.42,6.80       & 5.70,5.70,5.70        \\ % 10.08,10.08,8.88
Fiter, EPIC mode                & Thin, Full frame     & Medium, Full frame    \\
\hline
\end{tabular}
\label{tab:xmmlog}
\end{table*}
% % ---------------------------------------------------------------

%%%%%%%%%%%%% PROCESSINg
\subsection{Data Processing}\label{sec:data.processing}
\begin{figure}
\includegraphics[width=3.4in]{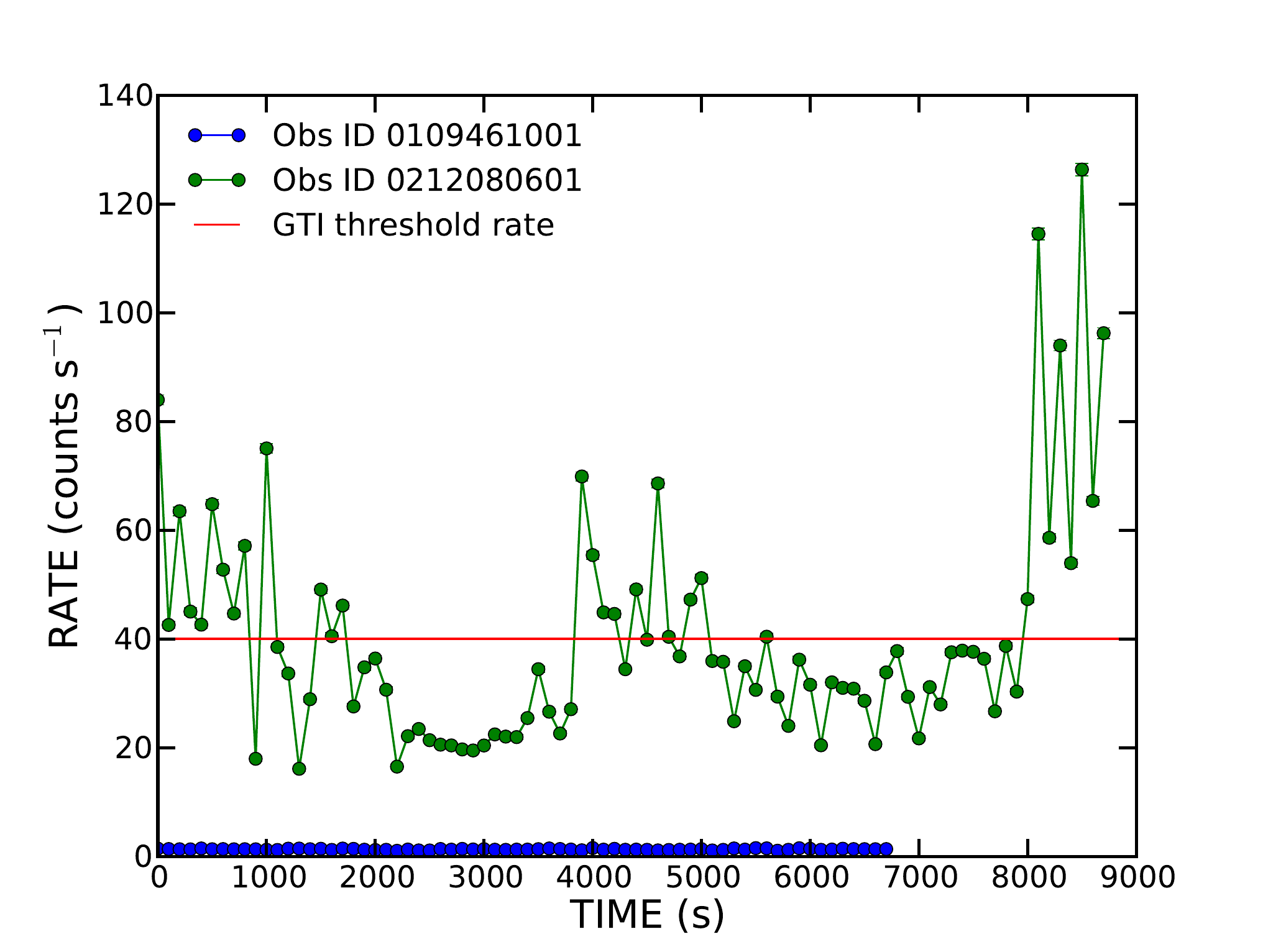}
\caption{The $>$10 keV PN light curves for the two XMM-Newton observations of M67, ID 0109461001 (blue) and ID 0212080601 (green). 
The latter has enormous soft proton flaring, and the count rate threshold of 40 counts per second used to define the good time interval 
is shown with the red line.}
\label{fig:gti}
\end{figure}

The raw data were processed with the SAS 13.5.0 and HEASoft 15.1 packages using the procedure outlined in the 
{\it XMM} ABC and {\it XMM} SAS guidebooks.
In brief, we (i) generated the calibration information file using {\tt cifbuild} task and downloaded the relevant calibration files 
from the {\it XMM} calibration archive, 
(ii) ran {\tt odfingest} to compile housekeeping information on the ODFs, 
(iii) applied the calibration with the {\tt emchain} and {\tt epchain} tasks, 
(iv) filtered the MOS and PN event lists by selecting only the good events ($XMMEA\_EM$ and $XMMEA\_EP$ respectively).
We then inspected the PN light curve for photons with energies $>$ 10 keV and found no evidence for soft-proton flaring 
in the ID 0109461001, but ID 0212080601 was contaminated by a significant flaring.
To mitigate the flaring background in ID 0212080601, we chose a good time interval (GTI) where the PN countrate for energies 
$>$10 keV was more than 40 counts s$^{-1}$.
The $>$10 keV PN light curves for ID 0109461001 and ID 0212080601 along with the GTI countrate threshold for the latter 
are shown in Figure~\ref{fig:gti}.   Using the GTI removes a large fraction of the 
major flaring events and reduces the background in the image by a factor of 
four. % as determined from the 'scale parameters' histograms of mosaic-T and mosaic-T-nogti images in DS9.
Note that the countrate threshold chosen for GTI is much larger than the standard 0.4 counts s$^{-1}$, but is acceptable for 
the detection of point sources. 

% ---------------------------------------------------------------
% FIGURE: GOOD TIME INTERVAL
% ---------------------------------------------------------------
% ---------------------------------------------------------------

%%%%%%%%%%%%% IMAGING
\subsection{X-ray Images}\label{sec:data.imaging}
We produced images in three energy bands, a soft ($S_X$) band (0.2 -- 0.5 keV), a medium ($M_X$) band (0.5 -- 2.0 keV) and a 
hard ($H_X$) band (2.0 -- 7.0 keV) band, 
for all three EPIC detectors using the {\tt evselect} and {\tt emosaic} tasks with the filtered MOS and PN event lists 
from \S\ref{sec:data.processing} as input.
Total energy band (0.2 -- 7.0 keV) images for each detector were also produced.
Although {\it XMM-Newton} is sensitive up to 10 keV, we restricted our images to $<$7 keV in order to 
enable easy comparison with the Chandra observation of M67 \citep{berg04}.
During the imaging process, only events with (i) the $FLAG$ keyword set to zero, (ii) $PATTERN$ keyword less 
than or equal to 4 for PN and (iii) $PATTERN$ keyword equal to zero for PN in the soft band were selected.
The motivation for the latter is mainly the rejection of noise at the extremities of the PN detector along the detector Y-direction.
This stricter requirement for $PATTERN=0$ for PN in the soft band is also implemented in the 2XMM catalog \citep{watson2009}, for example.
Image binning was so chosen as to produce 600$\times$600 pixel 4.35\arcsec pix$^{-1}$ images.

%%%%%%%%%%%%% SOURCE FINDING
\subsection{SOURCE DETECTION}\label{sec:data.source}
We used the {\tt edetect\_chain} script to simultaneously search for sources in the nine images (3 bands $\times$ 3 detectors) for each observation.
The subroutine {\tt edetect\_chain} calls a series of tasks sequentially: (i) {\tt expmap} to the calculate the exposure map for the input images 
using the attitude, vignetting, exposure, bad pixel information, (ii)  {\tt emask} produced a detector mask based on the exposure, 
(iii) {\tt eboxdetect} was used in local mode to search for sources above a maximum likelihood (ML) of 5 simultaneously in all the input 
images in 5$\times$5 pix boxes (pixels surrounding the boxes are used for background estimation), 
(iv) {\tt esplinemap} removed the sources detected by {\tt eboxdetect} from the input images to produce smoothed background maps 
through spline fits to the residual images, 
(v) {\tt eboxdetect} was used in the map mode to find sources similar to the earlier {\tt eboxdetect} call, except this time the background map was used, 
(vi) {\tt emldetect} used the source locations from map mode {\tt eboxdetect} to perform simultaneous ML point-spread function fits \citep{cruddace88} to all input images and 
determine source parameters such as total counts, countrate, hardness ratios, etc. For {\tt emldetect} we specified a ML detection threshold of six (similar to the 2XMM catalog; 
this corresponds to a 0.2\% probability for Poissonian noise fluctuation to have caused the observed source counts).
The hardness ratios (discussed in detail in \S4.1) were calculated for the detected sources based on their countrate in different bands, 
and these are especially useful for characterising weak sources for which spectral analysis was not possible.

We combined the single-band and single-detector images into a single mosaic image using the {\tt emosaic} task, for each observation.
We inspected the mosaics by eye and further used the hardness ratios of detected sources as well as the knowledge of their optical counterparts to reject 
one false positive in ID 0109461001 from the {\tt emldetect} source list.
Seventy two (72) sources are thus detected in ID 0109461001 and 32 in ID 0212080601.
Twenty nine sources are common between the two observations and three are unique detections in ID 0212080601.
The conversion of count rates to fluxes is described in \S\ref{sec:analysis}.
The minimum 0.2--7 keV flux of detected sources in the two observations is $1.6\times10^{-15}$ and $4.8\times10^{-15}$ erg cm$^{-2}$ s$^{-1}$ respectively.
These correspond to X-ray luminosities of $1.4\times10^{29}$ and $4.2\times10^{29}$ erg s$^{-1}$ respectively at the assumed distance of 850 pc to M67.
The images in the total energy band were subjected to {\tt esensmap} to find the median count rate corresponding to likelihood threshold of six.
Converting these to fluxes we get the median flux detection thresholds of $8.9\times10^{-15}$ and $3.0\times10^{-14}$ erg cm$^{-2}$ s$^{-1}$ 
in the 0.2--7.0 keV energy range for ID 0109461001 and ID 0212080601 respectively.
% These correspond to X-ray luminosities of $7.7\times10^{29}$ and $2.6\times10^{30}$ erg s$^{-1}$ at a distance of 850 pc.
% Note that the minimum flux thresholds of the two observations is $4.8\times10^{-15}$ and $3.0\times10^{-14}$ erg cm$^{-2}$ s$^{-1}$, which are 
% $7.7\times10^{29}$ and $2.6\times10^{30}$ erg s$^{-1}$ at the distance of M67.
Our final X-ray source catalog is given in Table~\ref{tab:xraysrc}, which gives source parameters for the PN detector with count rates 
and total counts in the 0.2 -- 7.0 keV band.
The 1$\sigma$ source position uncertainties range between 0.5\arcsec and 4.0\arcsec, the median uncertainty being 1.5\arcsec.
% Energy conversion factors (ECFs) for S,M,H from WebPIMMS are 1.805,1.992,0.444 (MOS, Thin); 11.236,7.142,1.241 (PN, Thin), 1.518, 1.979,1.631 (MOS, Medium), 
% and 9.416,7.148,1.281 (PN, Medium) assuming a power-law spectral model with photon index of 1.7 and 
% observed absorption $N_{\rm H}=3\times10^{20}~\mbox{cm}^{-2}$ as in 2XMM.
% 2-12keV band: 0.38850039 (MOS Thin),  0.39123631 (MOS Medium),  1.09349371 (PN Thin),  1.12930548 (PN Medium)
%Go to WebPIMMS, convert from XMM/PN Med Count Rate 5' region into FLUX. Input energy range 2-12keV, src countrate = 1, 
% Galactic NH = 3e20 (ref: http://xmmssc-www.star.le.ac.uk/Catalogue/2XMM/UserGuide_xmmcat.html#SelExp), 
% Model of source = powerlaw with photon index == 1.7. Then submit and find the flux as 8.855e-12 (not absorbed) => ECF = 1/8.855e-12 / 1e11 = 1.12930548
% 0.2-12keV band: 0.942 (MOS Thin), 3.489 (PN Thin)

% ---------------------------------------------------------------
% TABLE: ALL X-RAY SOURCES
% ---------------------------------------------------------------
% \newpage
\begin{table*}
\tiny
\caption{Summary of EPIC detections in the field of M67. For each source we give the XMM-Newton source ID (NX), 
coordinates ($\alpha_{\rm J2000}$, $\delta_{\rm J2000}$), uncertainty in the coordinates ($\delta_{\alpha, \beta}$), PN total counts (counts) and countrate (ctr) 
in the total energy band (0.2--7 keV), hardness ratios (HR), ROSAT and Chandra counterparts (RX and CX from Belloni et al. and van den Berg et al. respectively), 
the distance between the EPIC source position and the optical counterpart ($d_{\rm XO}$), membership in M67 (Y/N: Yes/No, from Yadav et al.; n: no from 
multiwavelength data) and comments. "N!" denotes sources that are not proper-motion members according to Yadav et al. and Vereshchagin et al., but 
show other evidences of membership (e.g. spectral fitting, literature), thus leading to their final classification as members of M67.
The superscript $^a$ on the source ID denotes that the corresponding counts, countrate, and hardness ratios are from the 
MOS1 detector due to the source being in one of the gaps of the PN CCD.}
\label{tab:xraysrc}
\begin{tabular}{rllrcccclllllll} % 15 cols
\hline \hline
(1) & (2)                                      & (3)                                               & (4)                      & (5)    & (6)             & \multicolumn{2}{c}{(7)}& (8) & (9)  & (10)  & (11)  & (12)          & (13) & (14)\\
NX & \multicolumn{1}{c}{$\alpha_{\rm J2000}$}  & \multicolumn{1}{c}{$\delta_{\rm J2000}$}          & $\delta_{\alpha, \beta}$ & counts & ctr             & \multicolumn{2}{c}{HR} & RX  & CX   & V     & $B-V$ & $d_{\rm XO}$  & Mem & Identification\\
\cline{7-8}
   & \multicolumn{1}{c}{($^h$,$^m$,$^s$)}      & \multicolumn{1}{c}{($^{\circ}$,\arcmin,\arcsec)}  & $(\arcsec)$              &        &  (c\,ks$^{-1}$) &  HR1 &  HR2            &     &      & (mag) & (mag) & (\arcsec)     &     & \\ 
\hline
\multicolumn{15}{c}{Sources Detected in Obs ID 0109461001}\\
\hline
1 & 08:51:23.22 & +11:48:25.9 & 0.4 & 291(19) & 51(3) & $0.63\pm0.05$ & $-0.84\pm0.04$ & 11 & 5 & 14.32 & 0.71 & 1.6 & Y & E1740 \\
2 & 08:51:38.40 & +11:49:05.4 & 0.5 & 286(18) & 53(3) & $0.59\pm0.06$ & $-0.65\pm0.05$ &  & 2 & 20.79 & 0.15 & 1.2 & n & E510; Gal,z$\sim$0.6 \\
3 & 08:50:57.09 & +11:46:07.6 & 0.5 & 208(19) & 94(8) & $0.39\pm0.09$ & $-0.50\pm0.09$ & 15 & 4 & 23.27 & 1.36 & 0.2 & n & SDSSJ0850.95; AGN2,z$\sim$0.5 \\
4 & 08:51:13.30 & +11:51:39.5 & 0.6 & 169(14) & 37(3) & $0.68\pm0.07$ & $-0.91\pm0.05$ & 8 & 1 & 13.57 & 1.01 & 1.6 & Y & E1708 \\
5 & 08:51:23.78 & +11:49:49.3 & 0.5 & 151(13) & 28(2) & $0.82\pm0.06$ & $-0.90\pm0.05$ & 10 & 6 & 11.44 & 0.90 & 0.1 & Y & Y1289 \\
6$^a$ & 08:51:18.64 & +11:47:02.5 & 0.6 & 125(12) & 14(1) & $0.70\pm0.08$ & $-0.74\pm0.07$ & 13 & 9 & 12.67 & 0.74 & 1.1 & Y & E1726 \\
7 & 08:51:28.30 & +11:41:27.4 & 0.7 & 123(12) & 26(2) & $0.55\pm0.09$ & $-0.56\pm0.10$ &  & 12 & 22.40 & 0.12 & 1.6 & N & E2024 \\ % AGN2
8 & 08:51:07.21 & +11:53:02.0 & 0.6 & 118(16) & 53(7) & $0.70\pm0.11$ & $-0.65\pm0.13$ & 7 & 10 & 12.68 & 0.69 & 0.7 & Y & E1687 \\
9 & 08:51:51.35 & +11:35:47.5 & 0.8 & 72(10) & 28(3) & $0.39\pm0.13$ & $-0.64\pm0.14$ & 59 &  & 20.99 & 0.66 & 2.0 & n & E1907; QSO1 \\
10 & 08:51:37.79 & +11:50:56.6 & 0.8 & 68(9) & 13(1) & $0.58\pm0.12$ & $-0.90\pm0.08$ & 40 & 16 & 13.64 & 0.45 & 1.4 & y & E1791; AH Cnc (F7V) \\
11 & 08:51:49.93 & +11:49:52.0 & 0.9 & 67(9) & 14(2) & $0.39\pm0.14$ & $-0.68\pm0.13$ & 43 & 20 & 20.98 & -0.05 & 0.9 & N & E2757 \\ % AGN1
12 & 08:50:52.70 & +11:47:44.4 & 1.1 & 62(9) & 19(2) & $0.22\pm0.16$ & $-0.40\pm0.17$ & 49 & 21 & 21.86 & 0.27 & 1.4 & n & E2183; AGN2 \\
13 & 08:51:33.02 & +11:34:11.9 & 1.3 & 62(9) & 26(4) & $0.74\pm0.11$ & $-0.77\pm0.13$ & 60 &  & 21.03 & 0.56 & 1.7 & n & E2593; Gal \\
14 & 08:51:04.52 & +11:55:28.1 & 0.8 & 61(9) & 20(3) & $0.72\pm0.13$ & $-0.45\pm0.14$ & 34 & 14 & 21.99 & 0.81 & 3.1 & n & E2823; Gal \\
15 & 08:50:37.97 & +11:47:09.3 & 1.2 & 54(9) & 24(3) & $0.72\pm0.16$ & $-0.18\pm0.17$ & 12 &  & 22.94 & 2.01 & 2.3 & n & SDSSJ0850.63; AGN2 \\
16 & 08:51:21.63 & +11:52:37.5 & 0.9 & 52(8) & 11(1) & $0.64\pm0.13$ & $-1.00\pm0.08$ & 37 & 24 & 11.26 & 0.66 & 1.9 & Y & Y1476 \\
17 & 08:52:16.79 & +11:48:31.0 & 1.1 & 50(8) & 20(3) & $0.69\pm0.13$ & $-0.83\pm0.12$ & 47 & 7 & 14.44 & 0.80 & 1.3 & Y & E1857 \\
18 & 08:51:03.86 & +11:46:29.4 & 1.6 & 50(11) & 20(4) & $-0.04\pm0.29$ & $0.05\pm0.30$ & 14 & 13 & 21.21 & 0.36 & 1.0 & N & E2153 \\
19$^a$ & 08:51:37.27 & +11:59:02.5 & 1.1 & 49(8) & 12(2) & $0.51\pm0.14$ & $-1.00\pm0.08$ & 31 &  & 11.62 & 0.06 & 1.3 & N & E1787 \\
20 & 08:51:28.12 & +11:49:26.6 & 0.9 & 48(8) & 8(1) & $0.52\pm0.15$ & $-0.93\pm0.10$ & 45 & 19 & 12.84 & 0.41 & 1.4 & Y & E1756 \\
21 & 08:51:04.76 & +11:45:56.3 & 1.1 & 47(8) & 12(2) & $0.87\pm0.10$ & $-1.00\pm0.06$ &  & 23 & 13.58 & 0.51 & 1.6 & Y & E1569 \\
22 & 08:51:17.71 & +11:44:30.4 & 1.4 & 43(7) & 8(1) & $0.71\pm0.13$ & $-0.99\pm0.05$ & 17 & 15 & 15.49 & 0.84 & 0.8 & Y & E1175 \\
23 & 08:51:46.08 & +11:49:49.9 & 0.9 & 42(7) & 9(1) & $0.62\pm0.16$ & $-0.67\pm0.16$ &  & 22 & 22.12 & 1.20 & 1.3 & n & E2756; Gal \\
24 & 08:51:50.15 & +11:46:05.9 & 1.4 & 41(7) & 8(1) & $0.61\pm0.15$ & $-1.00\pm0.10$ & 52 & 47 & 10.70 & 0.96 & 1.3 & N! & Y892 \\
25 & 08:52:13.52 & +11:46:21.1 & 1.7 & 39(8) & 13(2) & $0.72\pm0.16$ & $-0.81\pm0.16$ & 51 &  & 21.95 & 0.50 & 1.2 & N & E2147 \\ % AGN1
26 & 08:51:46.56 & +11:52:01.7 & 1.2 & 36(7) & 8(1) & $0.38\pm0.20$ & $-0.60\pm0.18$ &  & 33 & 20.74 & 0.18 & 1.4 & N & E600 \\ % AGN1
27 & 08:51:07.92 & +11:49:54.4 & 1.2 & 33(7) & 7(1) & $0.87\pm0.13$ & $-0.88\pm0.13$ & 42 & 17 & 15.68 & 0.86 & 2.5 & Y & E2759 \\
28 & 08:52:10.42 & +11:47:12.7 & 1.6 & 33(7) & 10(2) & $0.41\pm0.20$ & $-0.55\pm0.26$ &  &  & 22.08 & 0.10 & 2.4 & n & E2722; Gal \\
29 & 08:51:19.59 & +11:52:10.5 & 1.2 & 32(10) & 12(3) & $0.66\pm0.27$ & $-0.89\pm0.18$ & 38 & 48 & 14.00 & 0.60 & 2.0 & y & E1732; HX Cnc (G1V) \\
30 & 08:50:54.97 & +11:56:50.3 & 1.5 & 28(6) & 11(2) & $0.47\pm0.20$ & $-1.00\pm0.24$ &  & 58 & 15.90 & 1.04 & 0.7 & Y & E763 \\
31 & 08:52:07.40 & +11:52:07.1 & 1.4 & 27(6) & 8(2) & $-0.04\pm0.24$ & $-0.91\pm0.23$ &  &  & 21.30 & 0.77 & 0.6 & n & E2952; Fore* ($\sim$M7V) \\
32 & 08:51:35.76 & +11:47:59.7 & 1.1 & 25(6) & 4(1) & $0.77\pm0.31$ & $0.16\pm0.26$ & 48 & 31 & 22.38 & 0.34 & 1.7 & n & E2190; Gal \\
33 & 08:51:36.03 & +11:46:31.9 & 1.4 & 25(6) & 4(1) & $0.45\pm0.24$ & $-0.93\pm0.15$ & 50 & 49 & 12.79 & 0.63 & 2.0 & Y & E1784 \\
34 & 08:50:58.74 & +11:51:36.3 & 1.4 & 25(6) & 7(1) & $0.29\pm0.27$ & $-0.61\pm0.28$ & 39 & 32 & 20.07 & 0.23 & 3.1 & N & E1268 \\ % Gal
35 & 08:51:30.61 & +11:50:45.6 & 1.7 & 24(6) & 4(1) & $0.56\pm0.23$ & $-0.84\pm0.20$ &  & 29 &  &  &  &  &  \\
36 & 08:51:31.43 & +11:45:50.8 & 1.6 & 24(11) & 10(5) & $1.00\pm0.71$ & $0.14\pm0.46$ & 53 & 36 & 12.74 & 0.50 & 2.4 & Y & E1581 \\
37 & 08:51:20.83 & +11:53:27.2 & 1.7 & 24(9) & 13(5) & $0.83\pm0.31$ & $-0.62\pm0.34$ & 4 & 3 &  &  & 0.8 & N! & E3205 \\
38 & 08:51:52.93 & +11:36:54.7 & 1.5 & 24(6) & 8(2) & $0.59\pm0.22$ & $-0.55\pm0.26$ &  &  & 21.91 & 0.85 & 2.0 & N & E1929 \\ % AGN
39 & 08:52:07.44 & +11:41:13.2 & 1.2 & 24(6) & 8(2) & $-0.29\pm0.28$ & $-0.07\pm0.40$ & 55 &  & 21.52 & 0.15 & 1.8 & N & E2018 \\ % AGN1
40 & 08:51:11.60 & +11:50:29.5 & 1.5 & 23(6) & 5(1) & $1.00\pm0.15$ & $-0.37\pm0.24$ &  &  &  &  &  &  &  \\
41 & 08:51:30.47 & +11:33:39.8 & 1.8 & 23(6) & 10(2) & $0.41\pm0.23$ & $-0.89\pm0.25$ &  &  & 22.07 & 0.00 & 2.4 & n & E1866; AGN1 \\
42 & 08:51:27.27 & +11:46:55.1 & 1.5 & 23(6) & 3(1) & $-0.50\pm0.33$ & $0.23\pm0.46$ & 16 & 57 & 21.75 & 0.18 & 2.3 & N! & E2164, EU Cnc \\
43 & 08:51:04.66 & +11:42:19.3 & 2.2 & 22(6) & 6(1) & $0.53\pm0.24$ & $-0.91\pm0.18$ &  &  &  &  &  &  &  \\
44 & 08:51:27.13 & +11:45:02.4 & 1.3 & 22(7) & 5(1) & $0.64\pm0.26$ & $-0.87\pm0.29$ &  &  & 22.83 &  & 3.9 & n & E3674; Gal \\
45 & 08:51:24.69 & +11:48:59.8 & 2.4 & 22(7) & 3(1) & $0.56\pm0.29$ & $-0.67\pm0.27$ & 46 &  & 13.24 & 0.60 & 3.9 & Y & Y1212 \\
46 & 08:51:16.99 & +11:48:22.8 & 2.2 & 20(6) & 3(1) & $0.72\pm0.27$ & $-0.99\pm0.15$ &  &  &  &  &  &  &  \\
47 & 08:50:32.58 & +11:53:54.3 & 3.1 & 20(6) & 10(3) & $1.00\pm0.21$ & $-0.42\pm0.32$ &  &  &  &  &  &  &  \\
48 & 08:50:57.55 & +11:52:55.2 & 1.8 & 19(5) & 6(1) & $0.83\pm0.25$ & $-0.05\pm0.30$ & 36 &  &  &  &  &  &  \\
49 & 08:51:09.00 & +11:50:44.4 & 1.6 & 19(5) & 4(1) & $-0.44\pm0.80$ & $0.88\pm0.20$ &  & 39 & 20.62 & 1.40 & 1.6 & n & SDSSJ0851.14; Gal,z$\sim$0.4 \\
50 & 08:51:06.38 & +11:37:53.5 & 1.4 & 19(6) & 19(7) & $0.74\pm0.26$ & $-0.37\pm0.40$ &  &  & 20.36 & 0.27 & 2.3 & N & E111 \\ % AGN1
51 & 08:51:57.78 & +11:45:46.1 & 1.8 & 18(5) & 4(1) & $0.81\pm0.23$ & $-0.66\pm0.25$ &  &  & 15.59 & 0.71 & 1.6 & N & E1189 \\ % Fore*
52 & 08:50:53.19 & +11:59:24.6 & 2.6 & 17(5) & 9(3) & $0.35\pm0.41$ & $0.15\pm0.38$ &  &  & 22.17 & 0.00 & 2.6 & n & E2436; QSO1,z$\sim$1 \\
53 & 08:51:26.71 & +11:56:10.5 & 1.8 & 16(5) & 5(1) & $1.00\pm0.27$ & $-0.79\pm0.24$ &  & 61 & 15.90 & 1.07 & 2.0 & N! & E742 \\
54 & 08:51:27.02 & +11:55:23.9 & 1.5 & 15(5) & 4(1) & $1.00\pm0.29$ & $0.44\pm0.31$ & 32 & 8 & 23.46 & 2.36 & 1.2 & n & SDSSJ0851.45; AGN2,z$\sim$0.8 \\ % 75% prob in Yadav, ~K0V from fitting
55 & 08:51:44.01 & +11:48:49.7 & 1.8 & 15(5) & 3(1) & $0.93\pm0.17$ & $-0.77\pm0.29$ &  & 70 & 22.79 & 0.03 & 0.9 & n & E2208; Gal \\
56 & 08:51:36.30 & +11:44:43.8 & 2.3 & 15(5) & 3(1) & $0.56\pm0.33$ & $-1.00\pm0.32$ &  & 40 &  &  &  &  &  \\
57 & 08:52:05.18 & +11:50:42.9 & 2.4 & 14(5) & 4(1) & $1.00\pm0.21$ & $0.21\pm0.36$ &  &  &  &  &  &  &  \\
58 & 08:51:38.40 & +11:54:13.8 & 2.2 & 14(5) & 3(1) & $0.32\pm0.34$ & $-0.46\pm0.41$ &  & 71,149 & 21.42 & 0.57 & 0.9 & N & E2807 \\ % AGN
59 & 08:50:56.41 & +11:37:31.2 & 1.4 & 14(5) & 6(2) & $0.78\pm0.42$ & $0.28\pm0.34$ &  &  & 22.31 &  & 2.3 & n & E3586; QSO2,z$\sim$0.7 \\
60 & 08:51:47.73 & +11:39:54.6 & 3.3 & 14(5) & 4(1) & $-0.01\pm0.36$ & $-0.81\pm0.38$ &  &  &  &  &  &  &  \\
61 & 08:51:38.72 & +11:58:21.3 & 2.3 & 13(4) & 4(1) & $0.08\pm0.39$ & $-0.51\pm0.44$ &  &  &  &  &  &  &  \\
62 & 08:51:04.43 & +11:41:51.6 & 1.9 & 12(4) & 3(1) & $0.58\pm0.30$ & $-1.00\pm0.19$ & 54 &  & 19.36 & 1.66 & 2.5 & Y & E1498 \\
63 & 08:51:01.48 & +11:49:26.9 & 2.1 & 12(4) & 3(1) & $0.28\pm0.36$ & $-0.83\pm0.34$ & 44 & 68 & 14.82 & 0.72 & 1.9 & N & E520 \\
64 & 08:51:18.20 & +11:50:15.3 & 2.5 & 11(4) & 2(1) & $0.94\pm0.28$ & $-1.00\pm0.19$ & 41 & 104 & 14.41 & 0.61 & 4.8 & Y & E1723 \\ % Y1342
65 & 08:51:42.75 & +11:34:08.8 & 2.5 & 10(4) & 5(2) & $0.81\pm0.26$ & $-1.00\pm0.18$ &  &  &  &  & 1.6 & n & E3280; AGN2,z$\sim$0.3 \\
66$^a$ & 08:52:10.06 & +11:55:28.1 & 1.6 & 10(5) & 8(4) & $0.45\pm0.55$ & $0.03\pm0.59$ & 33 &  & 17.24 & 1.73 & 1.1 & n & E721; Fore* ($\sim$M7V) \\
67 & 08:51:25.87 & +11:59:33.2 & 1.7 & 9(4) & 5(2) & $1.00\pm1.38$ & $0.68\pm0.31$ &  &  &  &  &  & n & AGN2? \\
68 & 08:51:25.06 & +11:57:48.4 & 4.2 & 9(4) & 3(2) & $0.50\pm0.44$ & $-0.80\pm0.48$ &  &  &  &  &  &  &  \\
69$^a$ & 08:51:26.29 & +11:37:06.3 & 2.6 & 9(3) & 1(0) & $1.00\pm0.51$ & $0.46\pm0.40$ &  &  &  &  &  &  &  \\
70$^a$ & 08:50:56.52 & +11:45:21.9 & 2.8 & 7(4) & 1(0) & $0.54\pm0.57$ & $-0.34\pm0.61$ & 24 & 53 & 20.27 & 0.34 & 1.1 & N & E362 \\ % AGN1
71 & 08:52:06.51 & +11:53:24.4 & 2.9 & 4(3) & 1(1) & $1.00\pm0.29$ & $-1.00\pm0.45$ &  &  &  &  &  &  &  \\
72$^a$ & 08:52:17.03 & +11:41:11.3 & 2.9 & 4(3) & 1(1) & $1.00\pm0.75$ & $-0.37\pm0.65$ & 56 &  & 16.85 & 1.81 & 2.1 & N & E207 \\ 
% 73 & 08:50:46.13 & +11:41:21.4 & 2.3 & 1(2) & 0(1) & $-1.00\pm3.89$ & $1.00\pm1.40$ &  &  &  &  &  &  &  \\ % Fore M* 
\hline
\multicolumn{15}{c}{New Sources Detected in Obs ID 0212080601}\\
\hline
73 & 08:50:36.96 & +11:54:05.8 & 1.6 & 31(9) & 19(6)& $0.66\pm0.20$ & $-0.48\pm0.33$ & 35 &  & 14.49 & 0.78 & 1.2 & Y & E1475 \\ % 15; Y1589
74$^a$ & 08:51:28.80 & +11:48:07.9 & 1.7 & 22(7) & 4(1) & $1.00\pm0.13$ & $-0.74\pm0.28$ &   & 41 &  &  & & & \\ % 20
75 & 08:50:58.56 & +11:48:05.0 & 1.9 & 17(9) & 4(2) & $-0.26\pm0.59$& $-0.14\pm0.92$ &    &  & 15.81 & 0.80 & 2.2 & Y & E4630 \\ % 30; Y1096
\hline                                   
\end{tabular}
\end{table*}

% ---------------------------------------------------------------

%--------------------------------------------------------------------------------------------
% COUNTERPARTS AND SOURCE CLASSIFICATION
%--------------------------------------------------------------------------------------------
\section{IDENTIFICATION OF X-RAY SOURCES} \label{sec:identification}

In order to identify the X-ray sources found in both XMM-Newton observations of M67 with known classes of astronomical sources, 
we searched for optical and previously known X-ray counterparts. Information from SIMBAD and infrared data from WISE were used 
as necessary in order to accurately classify sources (especially in case of AGN or foreground stars).
All identifications were checked by eye.
Counterparts from \cite{yadav08} were used to establish membership probabilities. 
Below, we describe the identification process, and give simple estimates of chance identification 
and expected background X-ray sources.

%%%%%%%%%%%%% X-RAY COUNTERPARTS
\subsection{X-ray Counterparts from {\it ROSAT} and {\it Chandra}} \label{sec:identification.xray}
Previous observations with {\it ROSAT} \citep{belloni93,belloni98} and {\it Chandra} \citep{berg04} were with flux detection thresholds of 
$9\times10^{-15}$ and $2\times10^{-16}~\mbox{erg cm}^{-2}~\mbox{s}^{-1}$ respectively (see Figure~\ref{fig:fov} for coverage).
A comparison of our X-ray sources with those in the published catalogs, shows that, 
out of 61 X-ray sources detected by \cite{belloni98} with the {\it ROSAT} PSPC, 44 are in the {\it XMM} field of view (fov), 
of which, 40 have one {\it XMM} counterpart within the positional error.
\cite{berg04} detected 158 X-ray sources using the {\it Chandra} ACIS-I and ACIS-S detectors, of which 
153 Chandra sources lie within the {\it XMM} fov, and 41 of these have a probable {\it XMM} counterpart.
The X-ray counterparts are listed in column (8) and (9) of Table~\ref{tab:xraysrc}.
% The probability of spurious identification of these counterparts is analysed in Section~\ref{sec:identification.spurious}

%%%%%%%%%%%%% OPTICAL COUNTERPARTS
\subsection{Optical Counterparts} \label{sec:identification.optical}
The most comprehensive and sensitive survey of M67 is the ESO Imaging Survey Pre-FLAMES \citep[EIS; ][]{momany01}, 
which contains objects with $11\leq$V$\leq23$.
We cross-matched our X-ray source list with the EIS catalog to find optical counterparts.
The error in the optical source positions was taken to be $1\arcsec$, and this was added in quadrature to the X-ray source '
position uncertainties.  This procedure gave unique EIS counterparts for 50 XMM sources.
For sources absent in EIS, we searched for matches in the SDSS and \cite{yadav08} catalogs, thus finding 6 additional sources with optical counterparts.
In this paper we refer to the EIS and \citeauthor{yadav08} sources with the prefixes 'E' and 'Y' respectively, e.g. the optical counterparts of NX1 and NX5 are E1740 and Y1289 respectively.
The optical counterparts, their V magnitudes, $B-V$ colors and the distance between the X-ray source and the optical counterparts are 
listed in columns (14), (10), (11), and (12) of Table~\ref{tab:xraysrc} respectively.
Optical magnitudes and/or colors unavailable in the EIS or \citeauthor{yadav08} catalogs were estimated from other 
catalogs in VizieR Catalog Service\footnote{http://vizier.u-strasbg.fr/}, wherever applicable.
For SDSS counterparts, we converted u,g,r magnitudes to $V$ magnitudes and $B-V$ colors using the transformations for stars from \cite{jester2005}\footnote{http://www.sdss.org/dr4/algorithms/sdssUBVRITransform.html}.
Fourteen of our X-ray sources do not have an optical counterpart.
% Spurious identification of optical counterparts is dealt with in Section~\ref{sec:identification.spurious}.

%%%%%%%%%%%%% MEMBERS AND CLASSIFICATION
\subsection{Members of M67 and Source Classification} \label{sec:identification.mem}

There are several published proper motion membership studies for M67. 
\cite{sanders77,girard89,zhao93} used photographic plates and are limited to stars with V$\lesssim$16.
\cite{yadav08} calculated the proper motion membership probabilities for $\sim$2400 stars towards M67 having V$\lesssim$22.5.
% The precision of proper motions ranges from $\sim$2 mas yr$^{-1}$ at the bright end to $\sim$5 mas y$^{-1}$ at the faint end.
Recently, \cite{vereshchagin14} presented a revised list of M67 members among \citeauthor{yadav08} sources using the convergent point method.
We cross-matched our X-ray sources with \citeauthor{yadav08} catalog and found 38 matches.
For these sources, we obtained the membership information from \citeauthor{vereshchagin14}, thus finding 19 members and 19 non-members.
Among the sources lacking a counterpart in \citeauthor{yadav08}, we searched for membership information in \citeauthor{girard89} to 
get two additional members.
Note that for sources present in both, \citeauthor{yadav08} and \citeauthor{girard89}, we give precedence to the former.
For sources listed as non-members in \cite{vereshchagin14} that have $\geqslant75$\% probability in 
both, \citeauthor{yadav08} and \citeauthor{girard89} catalogs, we used the binarity and photometric information of the optical 
counterparts to argue membership, wherever applicable.
Two X-ray sources, NX24 (Y892) and NX42 (EU Cnc), found to be non-members in \cite{vereshchagin14} and \citeauthor{yadav08}, have been reclassified as members of M67 
based on archival photometric information and evidence from literature.
In total, we have detected 25 X-ray members of M67.

We classified the remaining 35 sources as AGN, quasar, galaxy, or foreground/background stars based on multiwavelength spectral energy distributions using 
photometry from the SDSS, WISE, 2MASS and NOMAD catalogs.
The membership information for sources having an optical counterpart is given in column (13) of Table~\ref{tab:xraysrc}.

Since the membership of X-ray sources found by \cite{belloni98} and \cite{berg04} was based primarily on the work by \cite{girard89}, we considered 
cross-matching their X-ray catalogs with \cite{yadav08}, and looking at the membership of the corresponding source as given by \cite{vereshchagin14}.
Among the X-ray sources from \citeauthor{belloni98} stated as non-members or sources without membership probability, we accept four (RX47=NX17=CX7; RX17=NX22=CX15; RX42=CX17; RX35=NX73) as members, 
and among their proposed members, we reject four (RX43=NX11=CX20; RX44=NX63=CX68 is a galaxy; RX49=NX12=CX21 is a type-2 AGN; RX19).
Note that CX20 (quasar) and CX68 were already classified as non-members by \cite{berg04}.
One source having unknown membership probability (RX54=NX62) in \citeauthor{belloni98}, has been classified as a member based on new proper motion information.
Similarly, RX23 is found to be a non-member.

Among the {\it Chandra} X-ray sources \citep{berg04} classified as probable non-members, we accept the following three 
as members based on their proper motion:
1. CX7=NX17=RX47 listed as a W UMa-type binary in SIMBAD; 
2. CX15=NX22=RX17 is a known binary with circularised 1.2-day orbit; and 
3. the unclassified source CX17=NX27=RX42).
Twelve sources close to the M67 main sequence were accepted as probable members by \citeauthor{berg04}.
Among these, we accept eight sources as true members of M67.
These are: 
(1) CX58=NX30 is listed as RS CVn in SIMBAD, and a known binary with a 3.6-day period, 
(2) CX61=NX53 with 75\% proper motion membership probability in \cite{yadav08}, and listed as a non-member in \cite{vereshchagin14}, 
is a W UMa-type binary with period 0.27 days, has a photometric parallax distance of $\sim$820 pc, 
and lies along the M67 main sequence, 
(3--5) CX62, CX77, and CX82 are members in \citeauthor{yadav08} and \citeauthor{vereshchagin14} with a high probability, 
(6--7) CX73, an M3V star, and CX76, a K7V star have estimated distances of $\sim$950 pc based on SDSS and WISE photometry. 
CX73 has a counterpart in \citeauthor{yadav08}, where it has a proper motion membership probability of 48\%, 
but it is listed as a non-member in \cite{vereshchagin14}.
However, the hardness ratios of this source in the Chandra source catalog \citep{evans2010} are consistent with that of an active star.
CX76 has no counterpart in \citeauthor{yadav08}, 
(8) CX80 is a member of M67 in \citeauthor{yadav08} and \citeauthor{vereshchagin14}.
Compared to the typical uncertainty in the {\it Chandra} X-ray source positions, the optical counterparts of CX73 and CX80 
are quite far away ($\sim$4\arcsec).
This could be attributed to the large off-axis distance of the location where these sources were detected on the ACIS CCD.
We reject the membership of the remaining four sources (CX117, CX129, CX141, and CX153) due to their very small 
membership probabilities in \citeauthor{yadav08}, and their classification as non-members in \citeauthor{vereshchagin14}.

\begin{figure}
\includegraphics[width=3.5in]{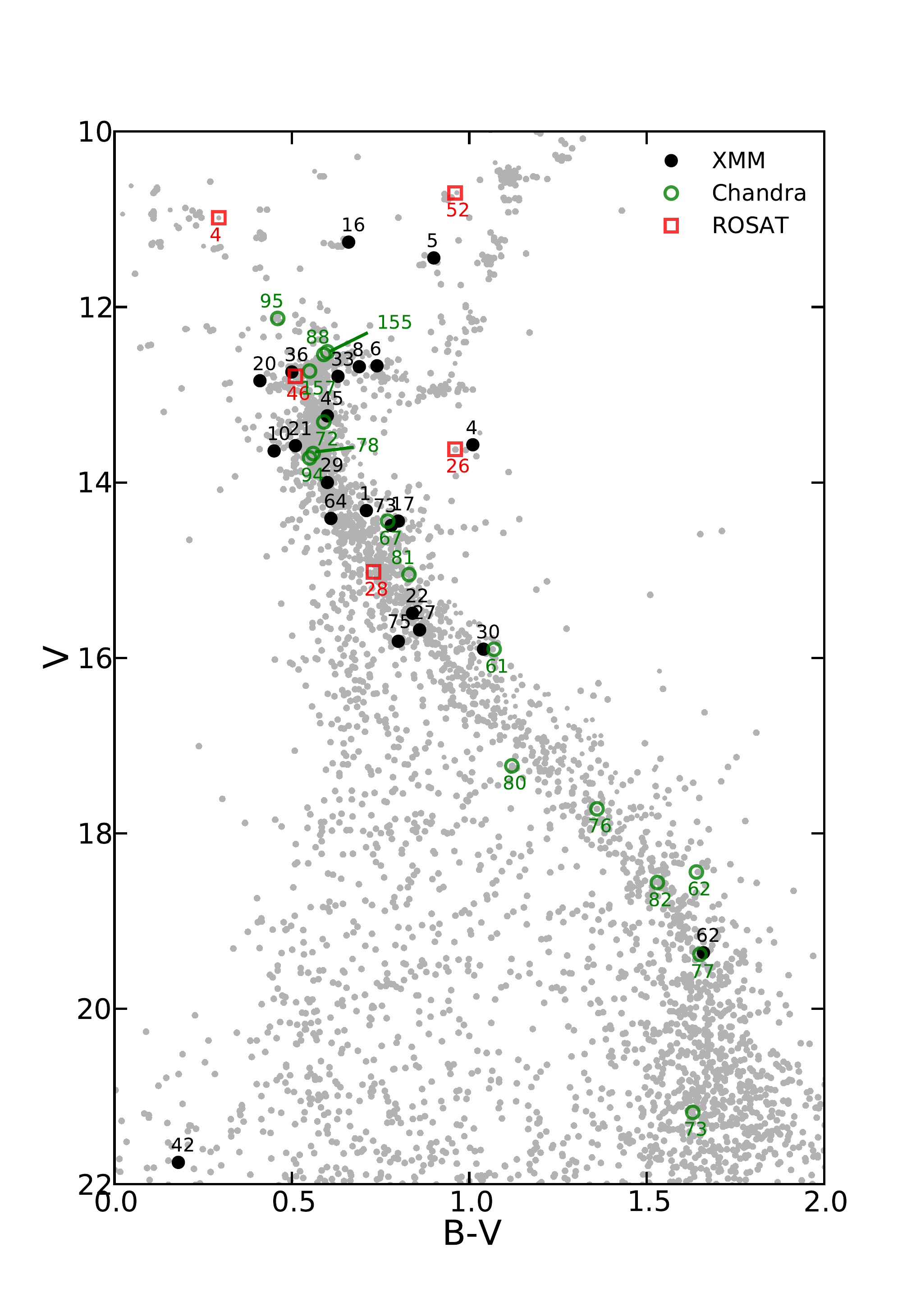}
\caption{$V/B-V$ color-magnitude diagram of X-ray members in M67 known till date, and listed in Table~\ref{tab:members}. 
Members from this work, from \citet{belloni98} and those from \citet{berg04} are plotted as black filled filled circles, red squares and green open circles respectively.
Numbers specified in these respective colors are the IDs of sources in the respective catalogs (NX, RX and CX).
Sources from WEBDA, %\footnote{http://www.univie.ac.at/webda}, 
from \citet{yadav08} with membership probability greater than 80\%, and all EIS sources are plotted in grey. 
Note that the $B-V$ colors for main sequence stars of spectral types A0, F0, G0, K0, M0, and M8 are 0.0, 0.3, 0.6, 0.8, 1.4, and 2.0 respectively \citep{johnson1966}.}
\label{fig:hr}
\end{figure}

A compilation of the spectral and orbital parameters and X-ray luminosities for all the 43 members of M67 detected 
as X-ray sources to date is given in Table~\ref{tab:members}.
Here, we list the optical counterparts, the spectral types \citep[from SIMBAD, VizieR, ][or estimated from archival 
multiwavelength photometry]{berg2000}, X-ray luminosity from 
our work or \cite{berg04} or \cite{belloni98} (in that order of preference), the orbital periods and ellipticities (from \cite{berg2000,berg04} or VizieR), 
and the source types (from \cite{berg04} or SIMBAD).
Note that the luminosities for sources detected in the {\it XMM-Newton} observations are quoted for the 0.2--7 keV energy band, and 
those from \citeauthor{berg04} are for the 0.3--7 keV energy band, while luminosities of \citeauthor{belloni98} sources were calculated using the 
{\it ROSAT} PSPC count rate converted to flux in the 0.2--7 keV energy band assuming an APEC 1.5 keV plasma model.
The orbital periods and eccentricities have been reproduced from \cite{berg2000,berg04}. 
For all known X-ray members of M67, we plot the $V/B-V$ color-magnitude diagram in Figure~\ref{fig:hr}.

% ---------------------------------------------------------------
% FIGURE: HR DIAGRAM
% ---------------------------------------------------------------
% ---------------------------------------------------------------

% ---------------------------------------------------------------
% TABLE: OPTICAL, CHANDRA AND ROSAT COUNTERPARTS
% ---------------------------------------------------------------
\begin{table}
\scriptsize
\caption{Summary of X-ray members of M67. The {\it top panel} shown members among the 
sources that were detected in XMM-Newton observations (this work) and classified as members in \citet{vereshchagin14}, 
The {\it middle panel} shows members detected in Chandra \citet{berg04}, while the {\it lower panel} lists 
members from ROSAT \citet{belloni98}. The source IDs from this work, from \citeauthor{belloni98} and \citeauthor{berg04} 
are given in columns 1--3, the optical counterpart and its spectral type are listed in columns 4--5, X-ray luminosity in column 6, 
the orbital period and ellipticity in columns 7--8, and source classification is in column 9. See \S\ref{sec:identification.mem} 
for details.}
\label{tab:members}
\begin{tabular}{p{0.1cm}p{0.1cm}p{0.1cm}lllp{0.2cm}p{0.2cm}l} % 9 cols
\hline \hline
(1)& (2)& (3)   & (4)    & (5)  & (6)                   & (7)           & (8)  & (9)\\
NX & RX & CX    & opt    & SpT  & $L_{\rm X}$/10$^{30}$ & $P_{\rm orb}$ & $e$  & comments \\
   &    &       &        &      & (erg s$^{-1}$)        & (d)\\ 
\hline
\multicolumn{9}{c}{Members from This Work}\\
\hline
1    & 11 & 5   & E1740  & G0V  & $6.7^{+1.1}_{-1.0}$   &  1.36         & 0    &  RS CVn\\
4    & 8  & 1   & HU Cnc & G8IV & $5.6^{+1.3}_{-1.1}$   & 18.40         & 0.21 &  SGB,RS CVn   \\
5    & 10 & 6   & Y1289  & G4III& $3.9^{+1.3}_{-2.1}$   & 42.83         & 0    &  III+WD,YS \\ % ref=van den Berg et al, 2000, ASP Conf Series
6    & 13 & 9   & HW Cnc & G4V  & $6.7^{+1.5}_{-1.4}$   & 10.06         & 0    &  RS CVn     \\
8    & 7  & 10  & HT Cnc & F9V  & $6.7^{+1.8}_{-1.7}$   &  1.36         & 0.10 &  RS CVn     \\
10   & 40 & 16  & AH Cnc & F7V  & $1.8\pm0.14$          &  0.36         &      &  Ecl W UMa\\
16   & 37 & 24  & Y1476  & G3III-IV & $1.5\pm0.14$      &  1495         & 0.32 &  SB,YS  \\
17   & 47 & 7   & LN Cnc &      & $2.7\pm0.4$           &  0.54         &      &  Ecl W UMa \\ % 39% Pmb in Girard % P_orb from Sandquist http://www.konkoly.hu/cgi-bin/IBVS?5679
20   & 45 & 19  & EV Cnc & F3   & $1.10\pm0.14$         &  0.44         &      &  W UMa     \\
21   &    & 23  & HS Cnc & F9V  & $1.65\pm0.27$         &  0.36         &      &  W UMa     \\
22   & 17 & 15  & E1175  & G9V  & $1.10\pm0.14$         &  1.17         & 0    &  RS CVn     \\ % 27% Pmb in Girard
27   & 42 & 17  & E2759  & G6V  & $0.96\pm0.14$         &               &      &  RS CVn?    \\
29   & 38 & 48  & HX Cnc & G1V  & $1.65\pm0.41$         &  2.66         & 0    &  RS CVn     \\
30   &    & 58  & HR Cnc &      & $1.51\pm0.27$         &  3.58         &      &  RS CVn     \\
33   & 50 & 49  & E1784  & G1V  & $0.55\pm0.14$         & 31.78         & 0.66 &  SB     \\
36   & 53 & 36  & E1581  & F9V  & $1.37\pm0.69$         &  4.36         & 0.0  &  RS CVn,triple \\
42   & 16 & 57  & EU Cnc & M5V  & $0.41\pm0.14$         &  0.09         & 0    &  CV, AM Her-type \\
45   & 46 &     & Y1212  & G0V  & $0.41\pm0.14$         &               &      &  \\
62   & 54 &     & E1498  &      & $0.41\pm0.14$         &               &      &  \\
64   & 41 & 104 & E1723  & F9V  & $0.27\pm0.14$         &   var         &      &  RS CVn?\\
73   & 35 &     & E1475  & G8V  & $3.06\pm0.97$         &               &      &  CV?\\
75   &    &     & E4630  & $\sim$K5V  & $0.65\pm0.32$   &               &      &  \\
\hline
\multicolumn{9}{c}{Members from \cite{belloni98}}\\
\hline
37   & 4  & 3   & ES Cnc & F5IV & $1.78\pm0.69$         & 1.07          &      & Algol \\ % Not a member from V, but 84% Pmb from Yadav (); 99% prob from Girard; ref for SpT=van den Berg & Verbunt 2000, ASP Conf Series
     & 26 &     & AG Cnc &      &     6.1               & 2.82          & 0    & RS CVn\\ % ref=van den Berg & Verbunt 2000, ASP Conf Series
     & 28 &     & HY Cnc & G6V  &     2.1               & 2.65          &      & RS CVn\\
     & 46 & 111 & E1589  & F7IV &     0.08              & 7.16          & 0    & RS CVn\\
24   & 52 & 47  & Y892   & G8III-IV & $1.10\pm0.14$     &  698          & 0.11 & RGB,YS \\ % not member in V but 96% Pmb from Yadav; 99% prob from Girard
\hline % mem from van den Berg 2004
\multicolumn{9}{c}{Members from \cite{berg04}}\\
\hline
53   &    & 61  & ET Cnc & $\sim$G4V  & $0.69\pm0.14$   & 0.27          &      & WU\\ % 75% prob in Yadav Y1740, not mem in V; not in Girard  % mem unknown in van den Berg
     &    & 62  & E683   &      &       0.40            &               &      & \\  % mem unknown in van den Berg
     &    & 67  & E1781  & G4V  &       0.36            &               &      & binary\\
     &    & 72  & E1677  & G0V  &       0.16            & 5.7           &      & RS CVn\\
     &    & 73  & E2650  & $\sim$M3V  & 0.46            &               &      & \\ % 48% membership in Yadav et al. and non-member in V; X-ray HR consistent with active corona form The Chandra Source Catalog, Release 1.1 (Evans+ 2012)
     &    & 76  & E1720  & $\sim$K7V  & 0.21            &               &      & binary\\ % absent in Yadav et al.
     &    & 77  & E394   &      &       0.23            &               &      & binary\\  % mem unknown in van den Berg; L_X estimated using rate, HR, and formula quoted in van den Berg et al. 2004
     &    & 78  & E1730  & F9V  &       0.30            & 5.95          & 0    & RS CVn\\
     &    & 80  & E2650  & $\sim$M3V  & 0.20            &               &      & \\ % member in Yadav et al. and V
     &    & 81  & E429   & G5V  &       0.12            & 6.7           &      & RS CVn\\  % 86% prob in Yadav Y980, not mem in V; mem in van den Berg
     &    & 82  & E1208  &      &       0.16            &               &      & \\  % mem unknown in van den Berg
     &    & 88  & E1729  &      &       0.09            & 7.65          & 0    & RS CVn\\
     &    & 94  & E1777  & F8V  &       0.16            &               &      & binary\\
     &    & 95  & E1590  & F6V  &       0.12            & 4913          & 0.34 & BS\\ % SpT ref: Liu et al. 2008, MNRAS, 390, 665
     &    & 155 & E2983  & F7V  &       0.04            & 11.02         & 0.26 & SB,RS CVn\\
     &    & 157 & HV Cnc & $\sim$F0V+M1V &       0.02   & 10.34         & 0    & Ecl Algol,triple\\ % 2013AJ....146..123G
\hline 
\multicolumn{9}{p{3.5in}}{Notes: (a) The countrates and hence the luminosity values from \cite{belloni98} and 
\cite{berg04} do not have associated uncertainties. (b) Some of the known binaries 
but do not have an orbital / period solution. (c) The orbital parameters are from \cite{mathieu1990}, \cite{latham1992}, \cite{berg2000}, 
and from unpublished work by D. Latham, R. Mathieu et al.}
% \multicolumn{9}{c}{Rejected Members from \cite{belloni98}}\\
% \hline
%      & 43 & 20  & \\
%      & 44 & 68  & \\
%      & 49 & 21  & \\ % Not a member from V, but 99% Pmb from Yadav; another possible counterpart not member in both
%      & 19 &     & \\% SDSSJ0849.9 & K3III 17\arcsec offset\\ %could be background AGN; Pasquini \& Belloni 1998 suggest uncertain identification
% \hline
\end{tabular}
\end{table}
% ---------------------------------------------------------------

%%%%%%%%%%%%% CHANCE IDENTIFICATION
\subsection{Chance Identification}\label{sec:identification.chance}
We estimated the probability of identifying an XMM--Newton source with an optical counterpart by chance using a procedure similar to the one adopted by \cite{berg04}.
This chance identification probability is a function of the positional uncertainty of the X-ray sources ($\Delta$) and the surface density of optical sources in the M67 field.
We calculated the mean positional uncertainty of the XMM--Newton sources, $\Delta=1.6$\arcsec, and the number of EIS sources inside the fov, N$\simeq$3035.
We used a circle with radius equal to the quadratic sum of $\Delta$ and the error in the optical positions, 1\arcsec, to calculate the search area around an X-ray source. 
Therefore, the probability that a randomly placed error circle includes an EIS-source is $N \pi (\Delta^2 + 1\arcsec^2) / A = 1.1 \times 10^{-2}$.
% $(\pi \times 1.9\arcsec^2) \times (2970/(\pi\times(16\times60\arcsec)^2))
Here, the area of the fov (red curve in Figure~\ref{fig:fov}) is denoted as $A$ approximated by a circle with radius 16\arcmin.
Thus, for the 75 X-ray sources that we have considered here (Table~\ref{tab:xraysrc}), the probability of 0, 1 or 2 chance identifications with EIS-sources 
% (75C$_r$ \times (1-1.6e-2)$^r$ \times 1.6e-2$^(n-r)$ where r is the number of chance identifications) 
is 42\%, 37\% and, 16\% respectively.
Similarly, we calculated the probability for the chance identification of our X-ray sources with M67 proper motion members to be 87\%, 12\%, and 1\% respectively.
We therefore conclude that one or two XMM--Newton sources have been falsely identified with optical counterparts while our M67 X-ray members are all likely to be genuine.
Another way of finding the chance identification probability is to note that among the 50 sources in \cite{berg04} having X-ray fluxes greater than 
the flux threshold of the XMM observations ($\sim$6$\times$10$^{-15}$ erg cm$^{-2}$ s$^{-1}$, corresponding to a {\it Chandra} count rate 
of 1 count ks$^{-1}$), about 40 have an optical counterpart.
We can then ask the question: {\it given the probability of finding an optical source counterpart for an X-ray source as 0.8, what is the probability of 
finding counterparts to 62 XMM sources out of 75?} Considering binomial distribution for the fraction of optical counterparts among X-ray sources, we get the answer as 10\%.
The most-likely value is 60 counterparts with a probability of 11.5\%.

We also estimated the probability for chance identification of {\it XMM--Newton} sources with {\it ROSAT} and {\it Chandra} sources.
There are 47 {\it ROSAT} sources in the {\it XMM--Newton} fov, having an average error radius (90\% probability) 
as 12.25\arcsec, so the chance identification of 0, 1, and 2 counterparts is 56\%, 33\%, and 9\% respectively.
For the {\it Chandra} sources we calculated the average error in RA and Dec separately, combined them with the corresponding 
average 1$\sigma$ positional error for {\it XMM--Newton} sources, and then used twice the resultant errors in 
RA and Dec as the sides of the error box.
Thus, considering the 154 {\it Chandra} sources in {\it XMM--Newton} fov we obtain the probability for 0, 1, and 2 chance 
identification as 97\%, 3\%, and 0\% respectively.
Accordingly, the probability of identifying {\it XMM-Newton} X-ray members with {\it ROSAT} and {\it Chandra} sources by chance is $<$1\%.

\subsection{Background sources} \label{sec:identification.background}

We used the soft AGN number counts from \cite{gilli2007} to estimate the expected background sources in our observations.
Using the flux limit of $5\times10^{-15}$ erg cm$^{-2}$ s$^{-1}$, we find from 
Figure 9 of \cite{gilli2007} that there are $\sim$230 AGN per deg$^2$ in the 0.5--2 keV band.
Then, among the 75 unique sources in the XMM fov (16\arcmin radius circle), we expect $\sim$50 AGN.
Among the 75 unique sources in our X-ray source list given in Table~\ref{tab:xraysrc}, 24 are probable M67 members (Table~\ref{tab:members}), two are foreground stars.
Four other non-members have V band magnitudes less than 17, and are probably foreground stars.
The remaining 45 sources have V$>$20 or have no optical counterpart down to $\sim$22nd magnitude. These are all very likely to be AGN.
In fact, we have confirmed that 18 of these are either galaxies (AGN status unknown) or AGN (see column 14 of Table~\ref{tab:xraysrc}).

%--------------------------------------------------------------------------------------------
% SPECTRA AND LIGHT CURVES
%--------------------------------------------------------------------------------------------
\section{SPECTRAL AND TIMING ANALYSIS} \label{sec:analysis}
\subsection{Spectral Fitting and Luminosities} \label{sec:analysis.spec}
We extracted PN background and source spectra for eight sources (NX1--8)  (Table~\ref{tab:xraysrc}) having total counts $>$ 100. 
The spectrum for one of the sources (NX6) had to be extracted from the MOS1 detector since it is located in the PN CCD gap.
To find the appropriate regions for the extraction of background and source spectra, we used DS9.
Spectral information was extracted from within a circle having 30\arcsec radius centred on the X-ray source. 
Background spectra were extracted from annuli with inner radii between 35\arcsec--45\arcsec and widths 
between 60\arcsec--75\arcsec centred on the X-ray sources, taking care to avoid contamination from nearby sources.
For those X-ray sources where the annulus would spill over to the neighboring CCD, the region of choice for background spectral extraction was a 
circle close to and having a similar RAWY as for the source extraction region, and with radius between 40\arcsec--50\arcsec.
We used the {\it SAS} task {\tt especget} to perform the extraction of source and background spectra for NX1--8.
We grouped the spectral (PHA) channels using the {\tt specgroup} task such that each grouped PHA channel had 16 counts, 
sufficient for the assumption of Gaussian distribution of uncertainties.
This task also performs the subtraction of background spectrum from the source spectrum.

% ---------------------------------------------------------------
% FIGURE: SPECTRA
% ---------------------------------------------------------------
\begin{figure*}
\includegraphics[width=1.92in,viewport=70 72 720 580,clip]{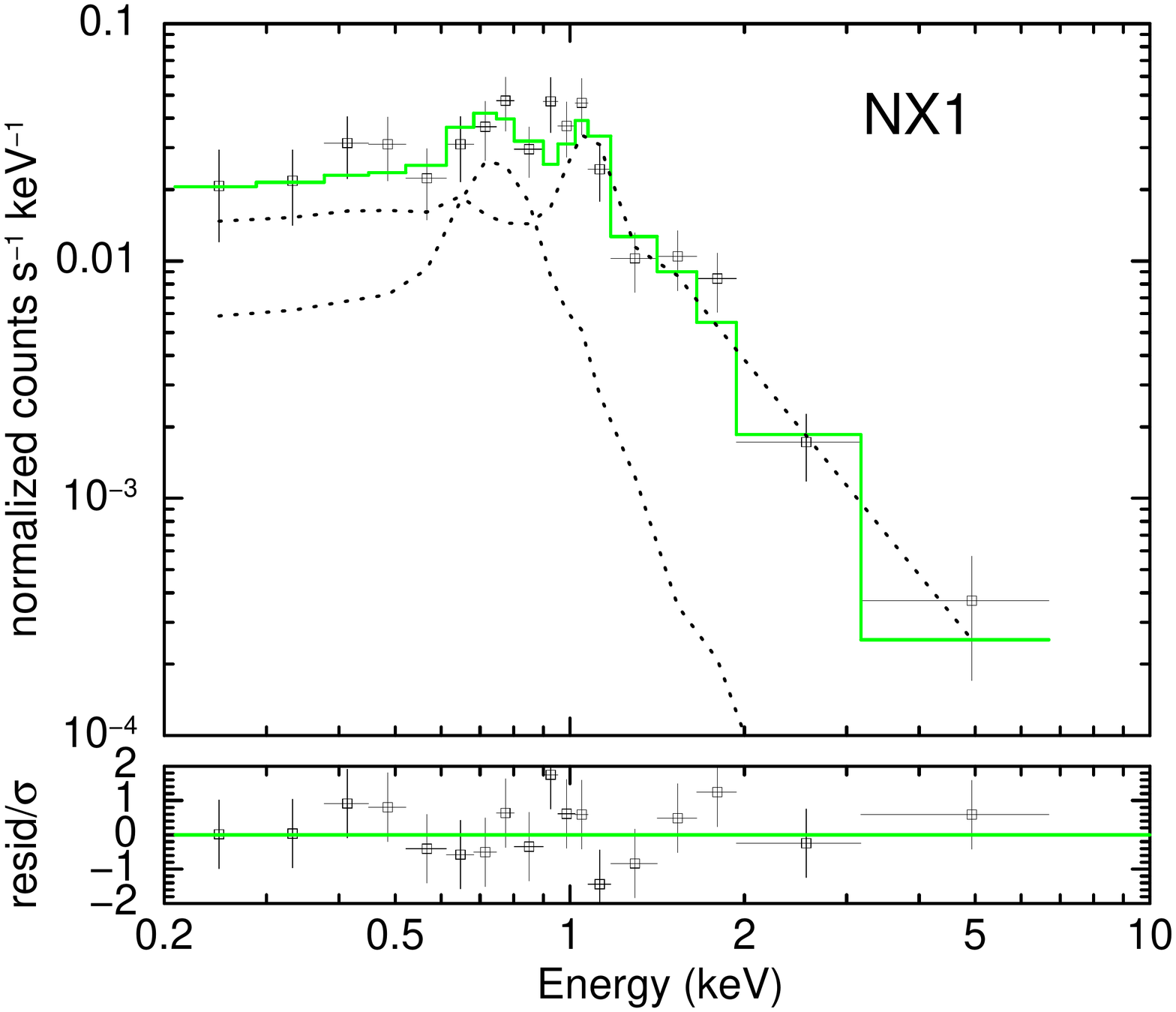}
\includegraphics[width=1.65in,viewport=160 72 720 580,clip]{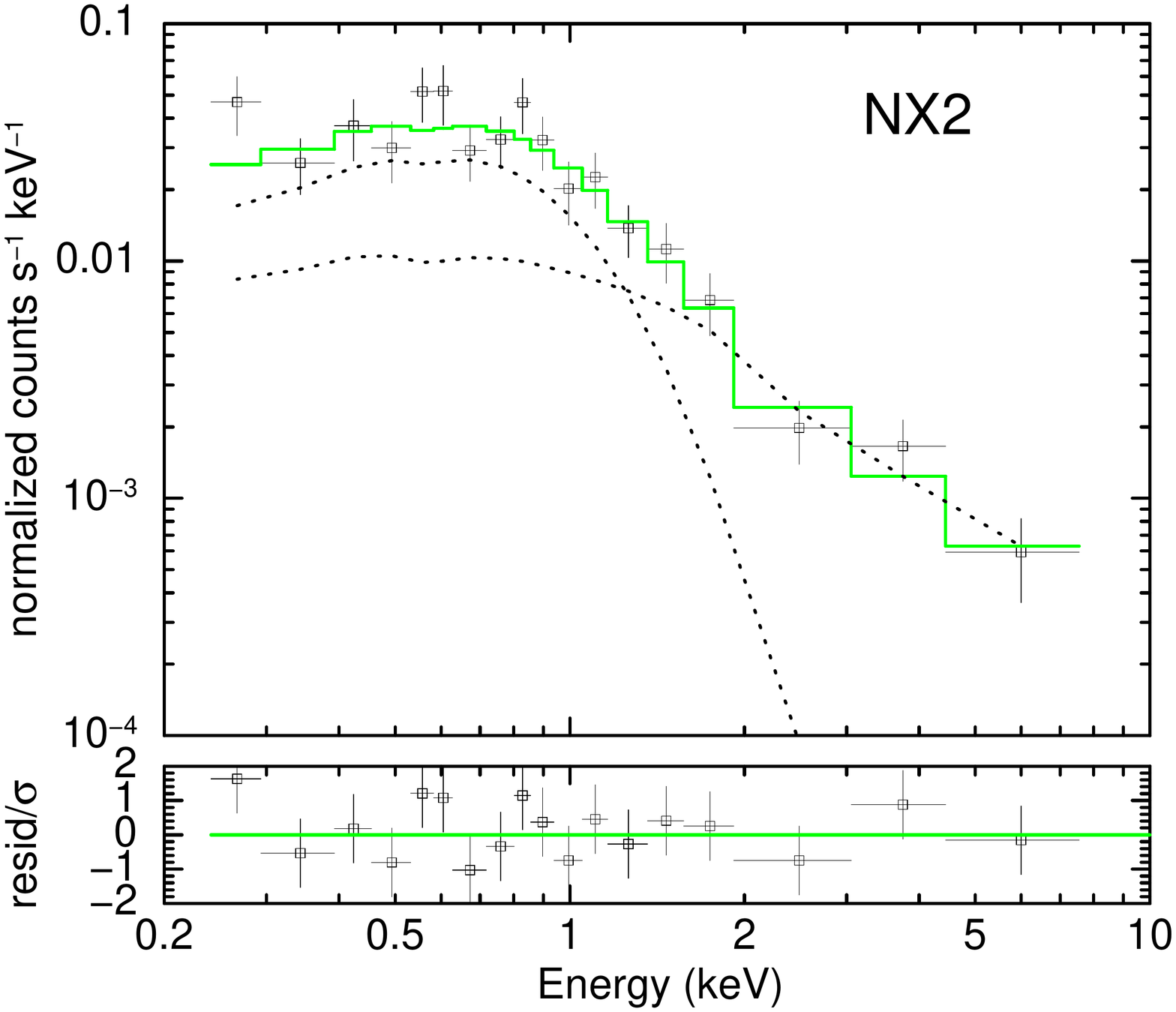}
\includegraphics[width=1.65in,viewport=160 72 720 580,clip]{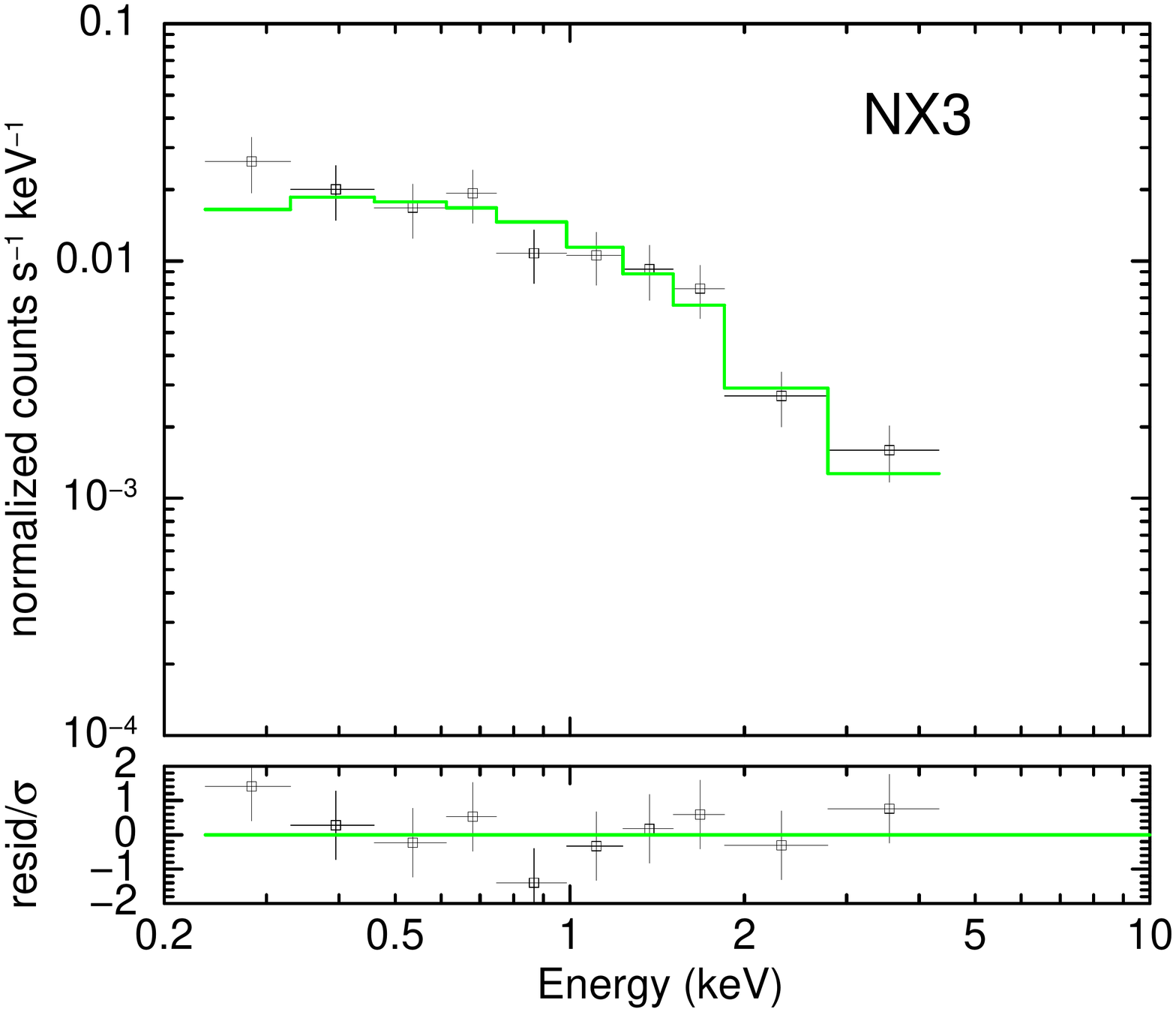}
\includegraphics[width=1.65in,viewport=160 72 720 580,clip]{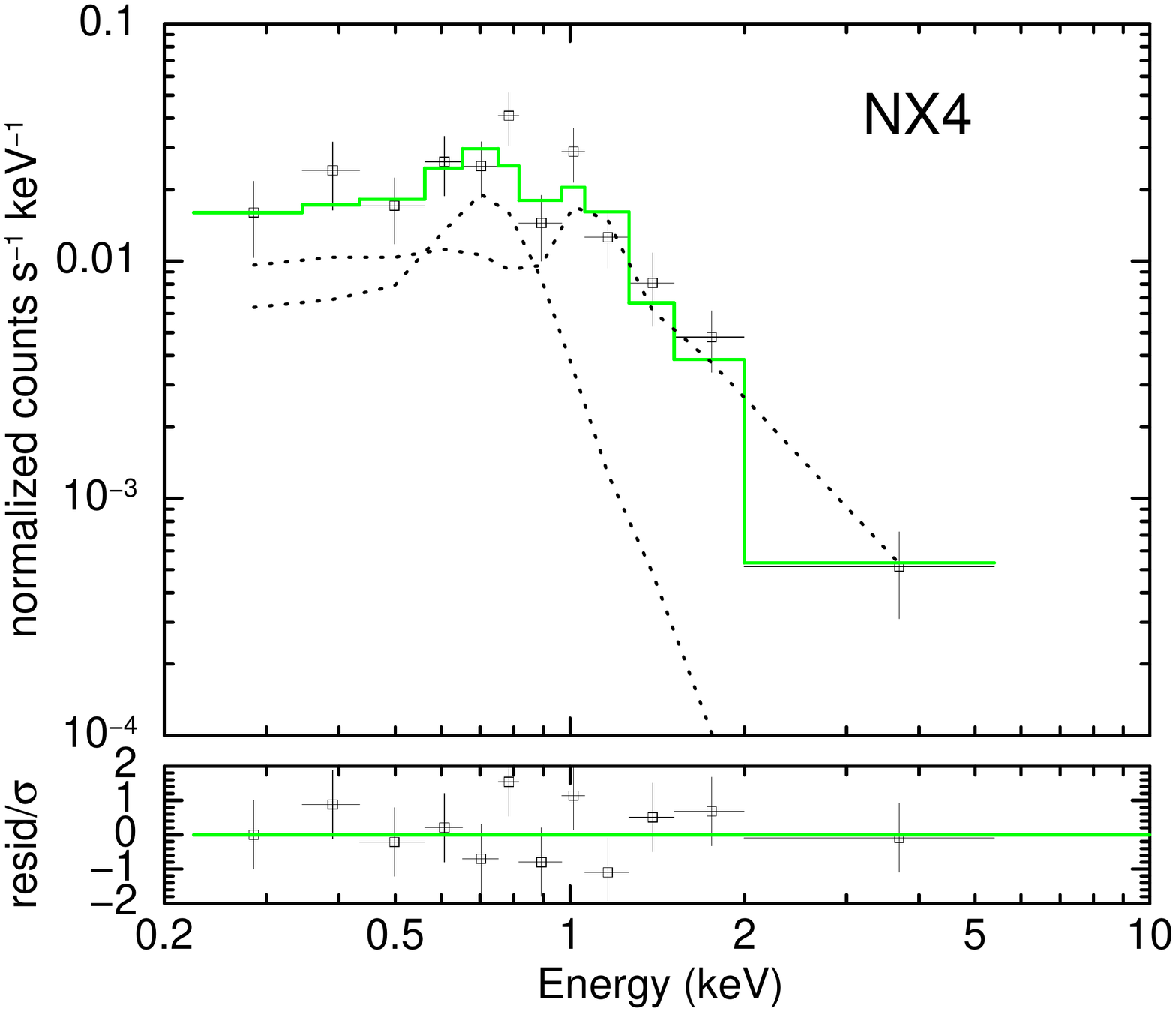} 
\\
\includegraphics[width=1.92in,viewport=70 12 720 580,clip]{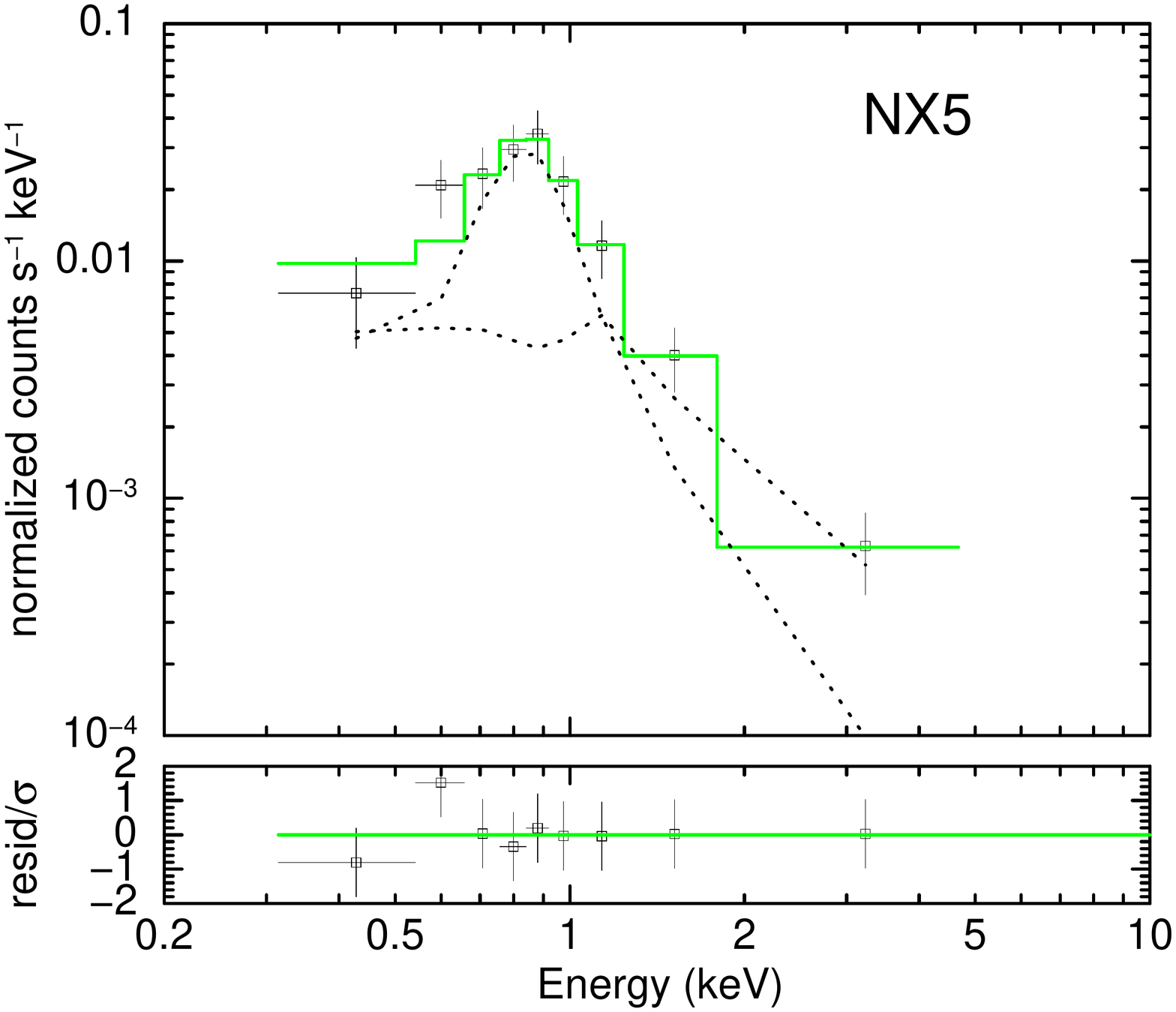}
\includegraphics[width=1.65in,viewport=160 12 720 580,clip]{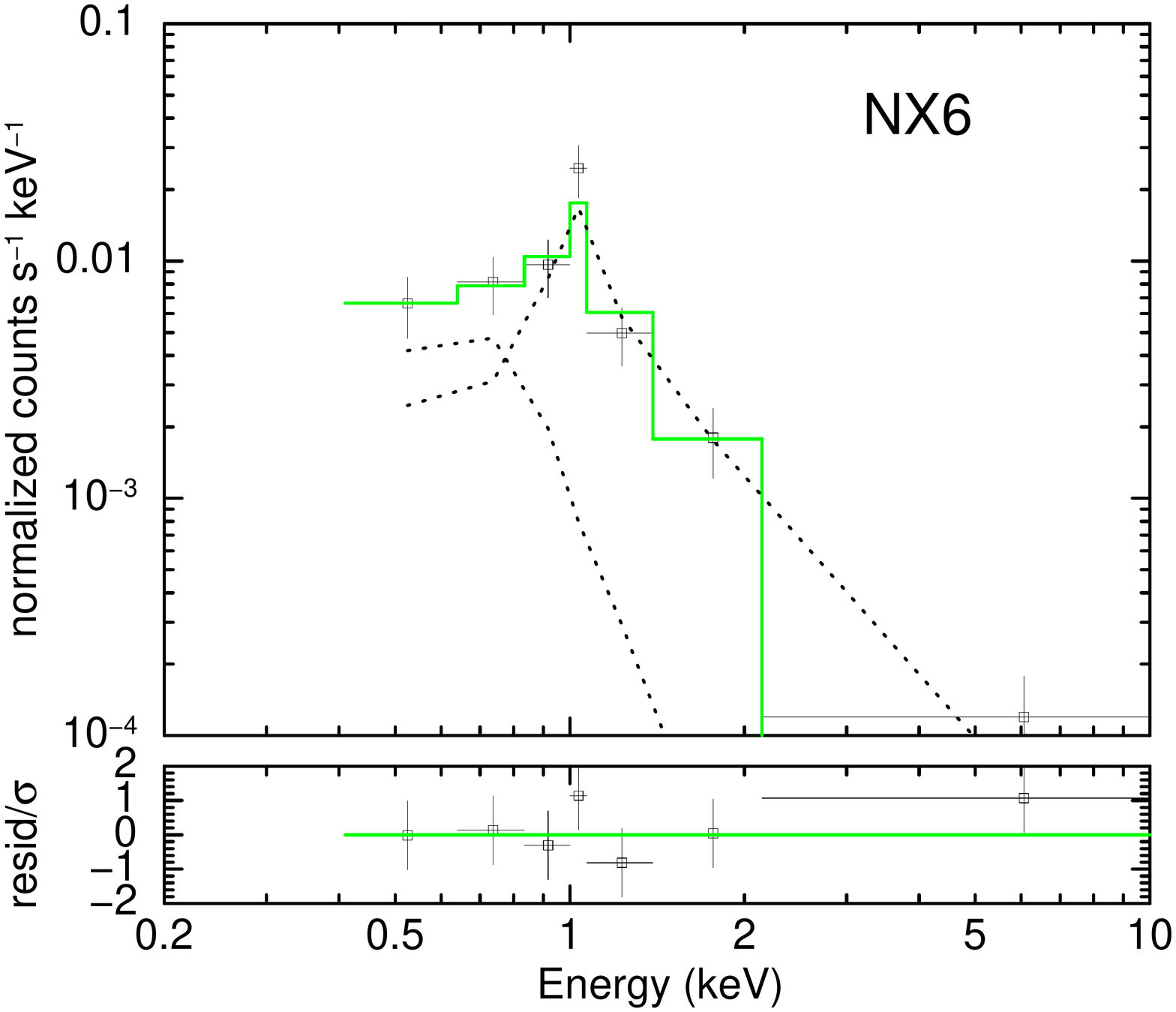}
\includegraphics[width=1.65in,viewport=160 12 720 580,clip]{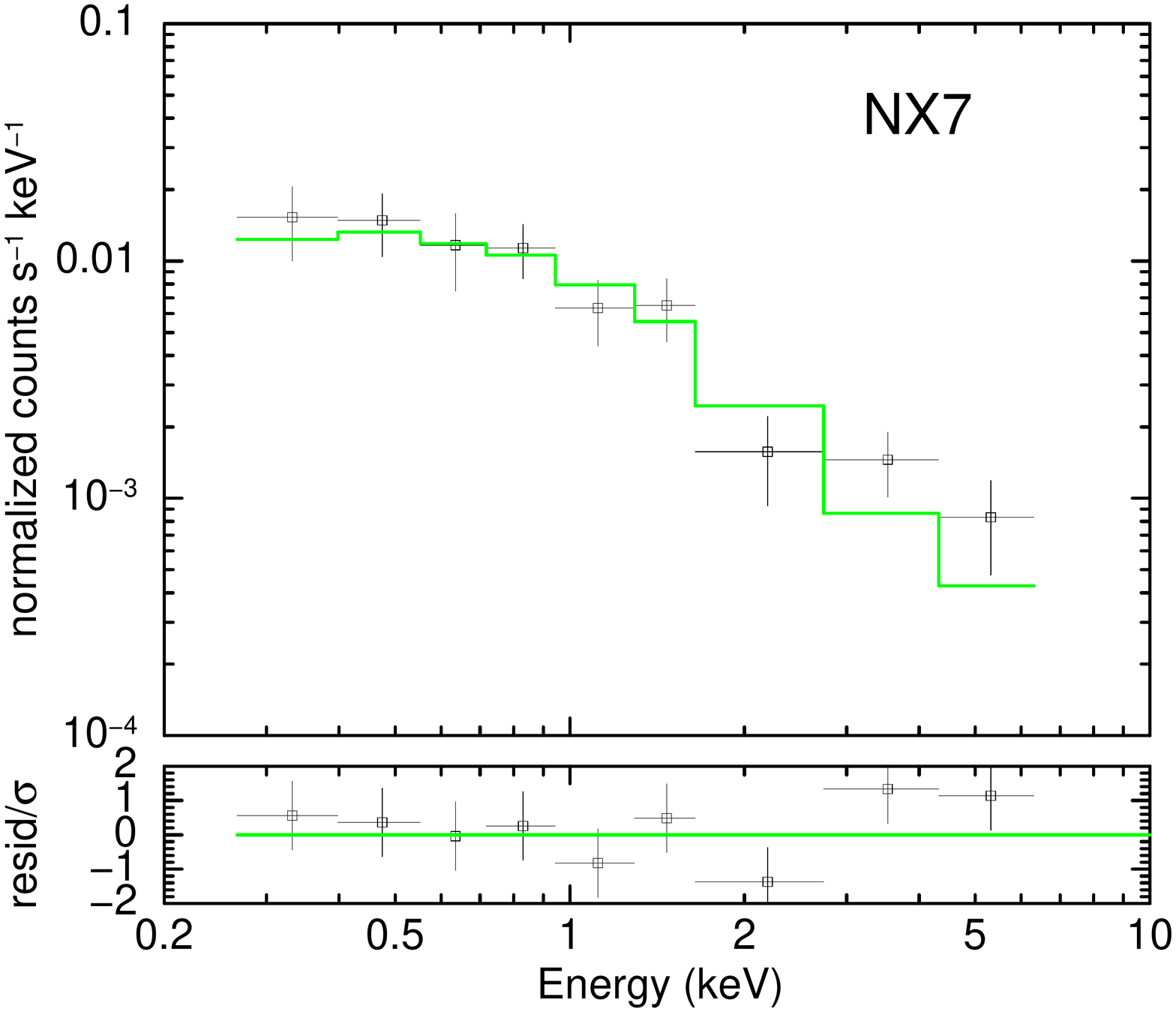}
\includegraphics[width=1.65in,viewport=160 12 720 580,clip]{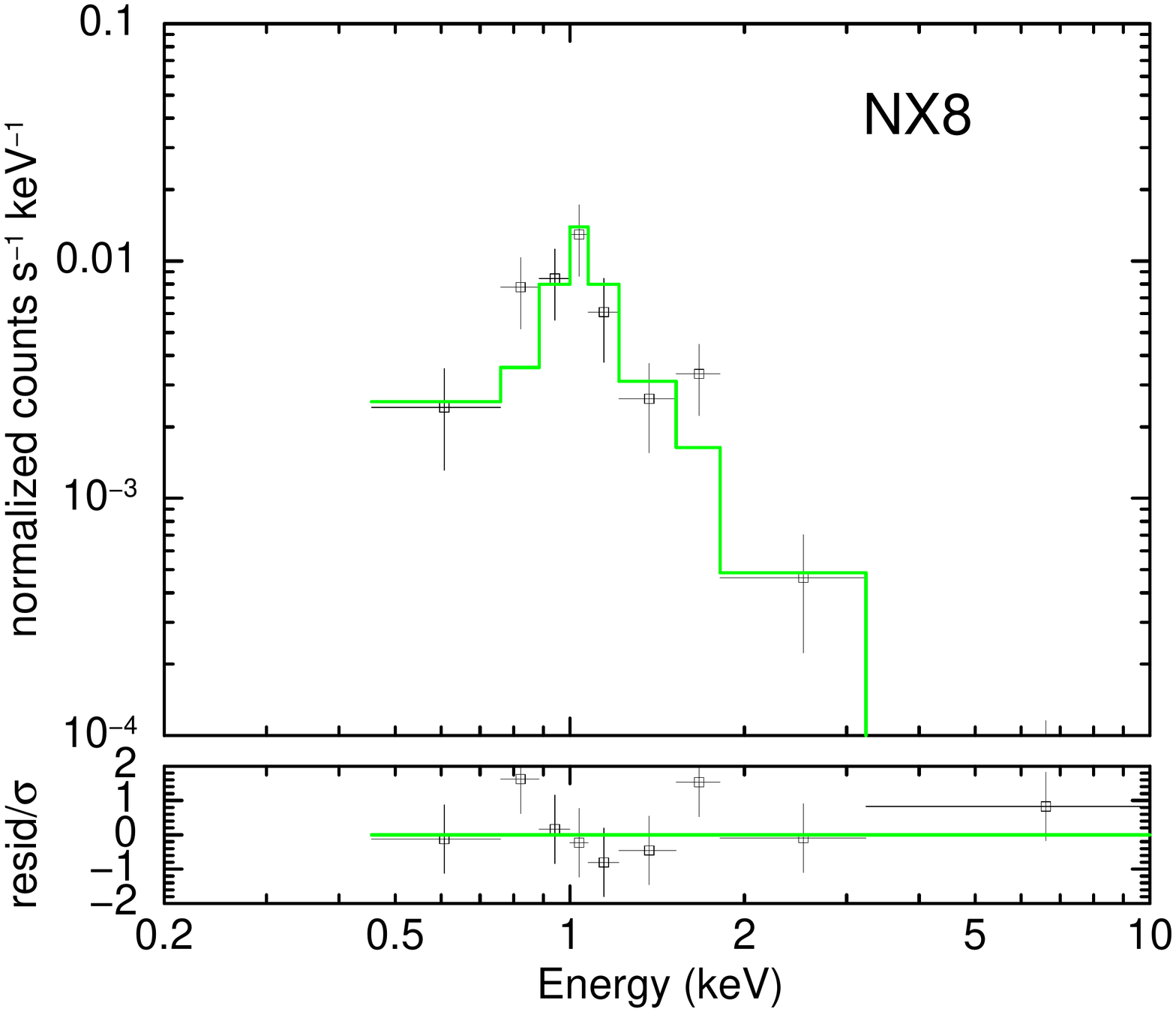}
\caption{Spectral fitting for sources having PN count rate greater than 100 (NX1--8). 
For each source, the top panel shows the binned background-subtracted normalised count rate (black points with error bars), 
the best-fit single-/multi-component spectrum (green), and the individual spectral components (dashed black curves), 
while the bottom panel shows the residual-to-noise ratio. 
For best-fit parameters, see Table~\ref{tab:spec}}
\label{fig:spec}
\end{figure*}
% ---------------------------------------------------------------

% ---------------------------------------------------------------
% TABLE: SPECTRAL FITTING
% ---------------------------------------------------------------
\begin{table*}
\caption{
The best-fit parameters and goodness-of-fit for the spectral fitting performed on sources NX1--8. The hydrogen column was held fixed and was calculated 
as described in \S\ref{sec:analysis.spec}. The columns are: (1) source ID; (2) single-multi-component model used for fitting along with photoelectric absorption; 
(3), (4) temperatures of the two APEC components used to fit spectra of member stars of M67, (5) ratio of the emission measures of the two APEC components; (6), (7) power-law index and 
blackbody temperature used for fitting non-members; (8) hydrogen column supplied to the photoelectric absorption model; (10) energy conversion factor (ECF=Rate/Flux; see \S\ref{sec:analysis.spec}); 
and (11) the reduced $\chi^2$ of the fit and the degrees of freedom (dof; equal to the number of channel groups minus one).
The corresponding fits are shown in Figure~\ref{fig:spec}. Note that NX6 fitting was done on MOS1.}
\label{tab:spec}
\begin{tabular}{lllllllllll} % 9 cols
\hline \hline
(1)& (2)   & (3)                    & (4)                    & (5)           & (6)                    & (7)                    & (8)                   & (9)                      & (10)                      & (11)\\
NX & model & kT$_1$                 & kT$_2$                 & EM1/EM2       & $\Gamma$               & kT                     & N$_H$                 & F/10$^{-14}$             & ECF/10$^{11}$             & $\chi_{\nu}^2$(dof)\\
   &       & (keV)                  & (keV)                  &               &                        & (keV)                  & ($10^{20}$ cm$^{-2}$) & (erg cm$^{-2}$ s$^{-1}$) & (cts cm$^{2}$ erg$^{-1}$) & \\
\hline
1  & 2T    & $0.39^{+0.66}_{-0.10}$ & $1.95^{+0.90}_{-0.48}$ & $0.17\pm0.10$ &                        &                        & 2.2                   & $7.73^{+1.22}_{-1.18}$   & 6.6                       & 0.82(14)\\
2  & P+B   &                        &                        &               & $1.36^{+0.62}_{-0.77}$ & $0.19^{+0.04}_{-0.04}$ & 3.3                   & $10.83^{+1.70}_{-2.85}$  & 4.9                       & 0.81(14)\\
3  & P     &                        &                        &               & $1.66^{+0.26}_{-0.25}$ &                        & 3.2                   & $19.95^{+4.89}_{-3.67}$  & 4.7                       & 0.69(8)\\
4  & 2T    & $0.32^{+0.20}_{-0.08}$ & $2.05^{+1.50}_{-0.51}$ & $0.24\pm0.14$ &                        &                        & 2.2                   & $6.45^{+1.50}_{-1.22}$   & 5.7                       & 0.95(8)\\
5  & 2T    & $0.76^{+0.16}_{-0.18}$ & $3.30^{+0}_{-1.86}$    & $0.46\pm0.29$ &                        &                        & 2.2                   & $4.55^{+1.54}_{-2.42}$   & 6.2                       & 0.63(5)\\
6  & 2T    & $0.24^{+0.24}_{-0.12}$ & $1.38^{+0.42}_{-0.15}$ & $0.43\pm0.30$ &                        &                        & 2.2                   & $7.71^{+1.73}_{-1.61}$   & 1.8                       & 1.08(3)\\
7  & P     &                        &                        &               & $1.71^{+0.38}_{-0.35}$ &                        & 3.3                   & $6.39^{+2.02}_{-1.41}$   & 4.1                       & 0.91(7)\\
8  & T     &                        & $1.51^{+0.55}_{-0.34}$ &               &                        &                        & 2.2                   & $7.71^{+2.06}_{-2.00}$   & 6.9                       & 0.94(7)\\
\hline
\end{tabular}
\end{table*}
% ---------------------------------------------------------------

The spectral fitting was done using the {\it XSPEC} package.
We ignored spectral groups containing channels with energies $<$ 0.2 keV and $>$12 keV.
X-ray spectra for the M67 members were fitted using an absorber model ({\tt phabs}) multiplied to a model of collisionally-ionised plasma 
({\tt apec} with one or two characteristic plasma temperatures and and both having default elemental abundances). The column density 
N$_{\rm H}$ for the absorber was fixed at 2.2$\times$10$^{20}$ cm$^{-2}$ (converted from E$_{B-V}$=0.04 using R$_V$=3.1).
Note that the choice of two {\tt apec} components instead of a single one was decided based on the reduced $\chi^2$ of the fit and 
manual inspection of the fitted spectrum and residuals.
For all the other X-ray sources, a photo-electrically absorbed power law with or without a blackbody component, 
as expected for AGN, was used for spectral fitting.
In this case, the N$_{\rm H}$ was estimated using the {\tt nh} task.
For NX1--8, the fitted model, their best-fit parameters with 90\% uncertainties given by {\it XSPEC}, and the N$_{\rm H}$ 
used are shown in columns 2--8 in Table~\ref{tab:spec}, and the resulting reduced $\chi^2$ and degrees of freedom (dof) 
are in column 11.  Figure~\ref{fig:spec} shows the spectra for the eight sources considered for fitting, of which five are members of M67.

The fluxes of NX1--8 in the 0.2--7 keV energy band were calculated using the best fit parameters in the {\tt flux} task in {\it XSPEC}.
The resulting fluxes along with their 90\% uncertainties are given in column 9 of Table~\ref{tab:spec}.
The energy conversion factors (ECF=Rate/Flux) are tabulated in column 10.
For members of M67 among sources in Table~\ref{tab:spec}, we find that the mean ECF is 6.3$\pm$0.4$\times10^{11}$ counts cm$^{2}$ erg$^{-1}$ (uncertainty in the count rate not folded in).
We used this factor to calculate the X-ray fluxes in the 0.2--7 keV band for all the M67 members detected in the {\it XMM} observations.
Similarly we determined the ECFs for the 0.3--7 keV and 0.1--2.4 keV bands.
The former band is used by \cite{berg04} for {\it Chandra} observations of M67 while the latter is relevant for {\it ROSAT}.
For coronal sources, we used the ECFs to calculate the multiplicative factors for converting from the {\it Chandra} and {\it ROSAT} 
bands to the 0.2--7 keV band as 1.0$\pm$0.1 and 0.8$\pm$0.1 respectively.
X-ray luminosities in the 0.2 -- 7.0 keV energy band of {\it XMM} for all the members of M67 calculated assuming a mean distance 
of 850 pc are given in Table~\ref{tab:members}.
For Obs ID 0212080601, we fitted the spectra for four bright members of M67 (NX4--6,NX8) using the parameters listed in Table~\ref{tab:spec} as initial values, and calculated the ECFs. 
The mean ECF was found to be $6.2\times10^{11}$ counts cm$^{2}$ erg$^{-1}$. 
We used this ECF to convert the count rates of NX73--75 to their corresponding fluxes.

% ---------------------------------------------------------------
% FIGURE: HARDNESS RATIOS
% ---------------------------------------------------------------
\begin{figure*}
\includegraphics[width=6in]{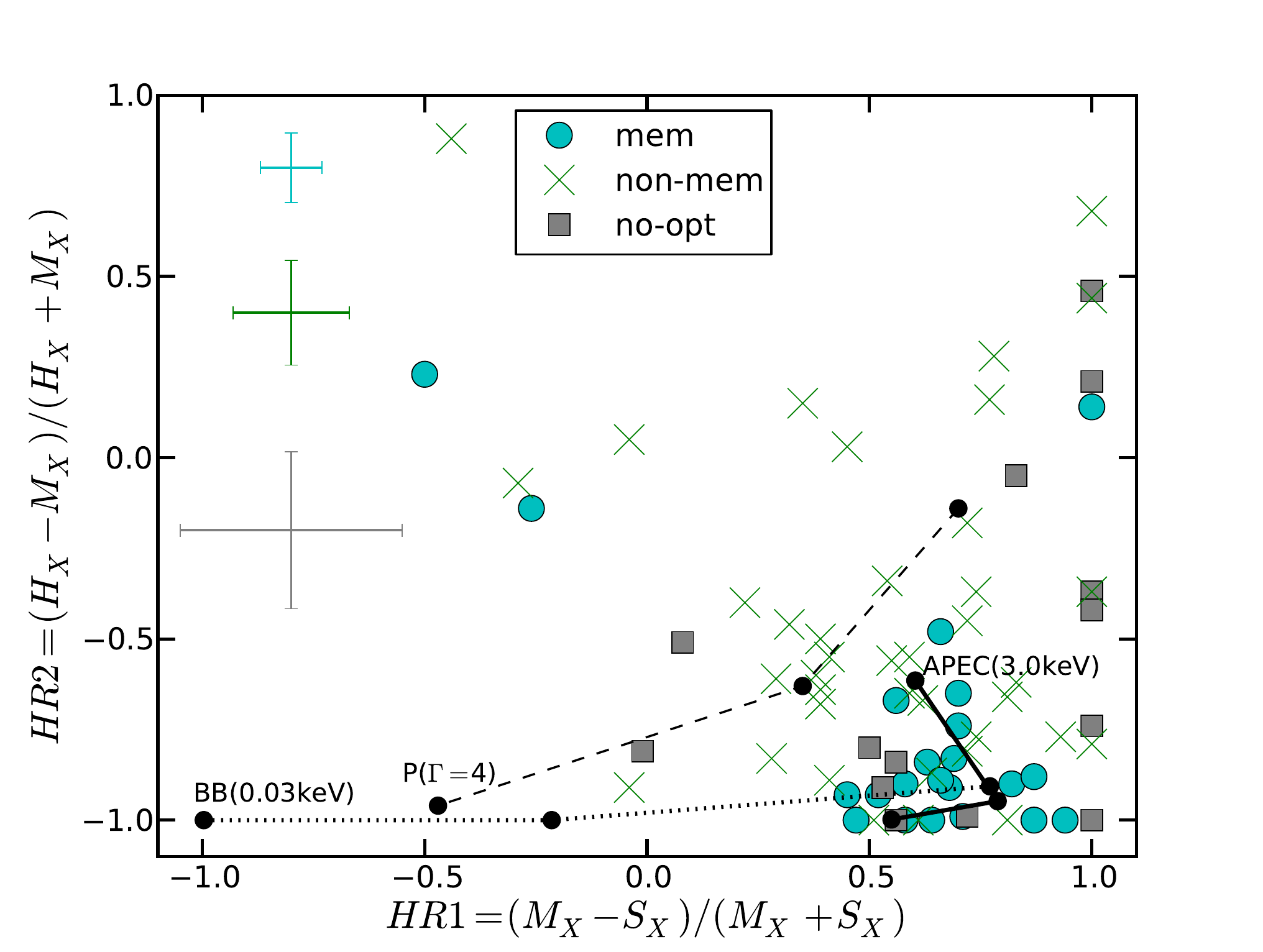}
\caption{Hardness ratios (see \S\ref{sec:analysis.hardness}) of the 75 unique sources detected in the XMM observations. Sources identified as M67 members, non-members, and 
sources with no optical counterparts (also suggested as non-members in \S\ref{sec:identification.mem}) are shown as cyan circles, green crosses, and grey squares respectively. 
The weighted mean error bars are shown to the left hand side of the plot, representative of sources in each of the three classes. Note that these error bars are just for guiding the eye, and 
the actual uncertainties (see Table~\ref{tab:xraysrc}) depend on countrate. Note also that the members NX36 and NX75 have large uncertainties in their hardness ratios ($\sim$0.6), and 
appear to be outliers with respect to the clustering of M67 members in the lower right hand corner of the plot denoting coronal emitters. 
The cataclysmic variable EU Cnc is located at $HR1=-0.50(\pm0.33)$, $HR2=0.23(\pm0.46)$.
The expected hardness ratios for three kinds of spectra, power-law (P($\Gamma$)), 
blackbody (BB), and collisionally-ionized plasma (APEC), are shown as black circles joined by black lines (dashed, dotted and solid lines respectively). 
Each black dot marks a unique value of temperature (in case of BB or APEC) or a power law index. the hardness ratios have been calculated for $\Gamma$=1, 2, and 4 in the case of power law; 
T=0.03, 0.1, and 0.3 keV for BB; and T=0.3, 1, and 3 keV for APEC. For each type of spectrum, one point has been labeled.
These were calculated using WebPIMMS using Galactic hydrogen column density of 2.2$\times$10$^{20}$ cm$^{-2}$.}
\label{fig:hardness}
\end{figure*}
% ---------------------------------------------------------------

%%%%%%%%%%%%% HR
\subsection{Hardness Ratios} \label{sec:analysis.hardness}
Spectral information for weak X-ray sources can be obtained by comparing the countrates in coarse spectral bins after 
applying the correction for exposure, vignetting, and background.
Taking advantage of the wide bandpass of {\it XMM}, we defined two hardness ratios, $HR1$ and $HR2$, as given in equation~\ref{eqn:hardness}.

\begin{equation}
HR1= \frac{(M_X - S_X)}{(M_X + S_X)}~ ,\qquad HR2= \frac{(H_X - M_X)}{(H_X + M_X)}
\label{eqn:hardness}
\end{equation}

\noindent where $S_X$, $M_X$ and $H_X$  denote the exposure-corrected counts in the soft band (0.2 -- 0.5 keV), 
the medium band (0.5 -- 2.0 keV) and the hard band (2.0 -- 7.0 keV), respectively.
For each X-ray source detected, HR1 and HR2 are calculated within the best-fit source position, using the best-fit count rates in all the bands 
by the source-finding algorithm {\tt emldetect}. They are tabulated in column 7 of Table~\ref{tab:xraysrc} for the PN detector (unless noted as MOS1).
Sources with very soft spectra will have HR1$<$0 and those with hard spectra will have HR2$>$0.
In Figure~\ref{fig:hardness} we plot HR2 against HR1 for all of our 75 unique sources.
Apart from NX36 and NX75, whose hardness ratios are imprecise, and the cataclysmic variable NX42=EU Cnc, all members of M67 lie in the region HR1$\gtrsim$0.5, HR2$\lesssim$0.5.
AGN and other non-members also tend to lie in this region, but their scatter is much larger.
Hardness ratios expected from power-law (P), blackbody (BB) and collisionally ionised plasma (APEC) are also shown.
These were calculated using WebPIMMS using Galactic hydrogen column density of 2.2$\times$10$^{20}$ cm$^{-2}$.
Note that the uncertainties of the source hardness ratios depend on the count rate and for our {\it XMM} sources they have a large scatter.
Only the weighted mean uncertainties are shown as guidance in Figure~\ref{fig:hardness} for each class of object.

%%%%%%%%%%%%% VARIABILITY
\subsection{X-ray Variability} \label{sec:analyses.variability}
\begin{figure*}
\includegraphics[width=1.76in,viewport=0 25 530 410,clip]{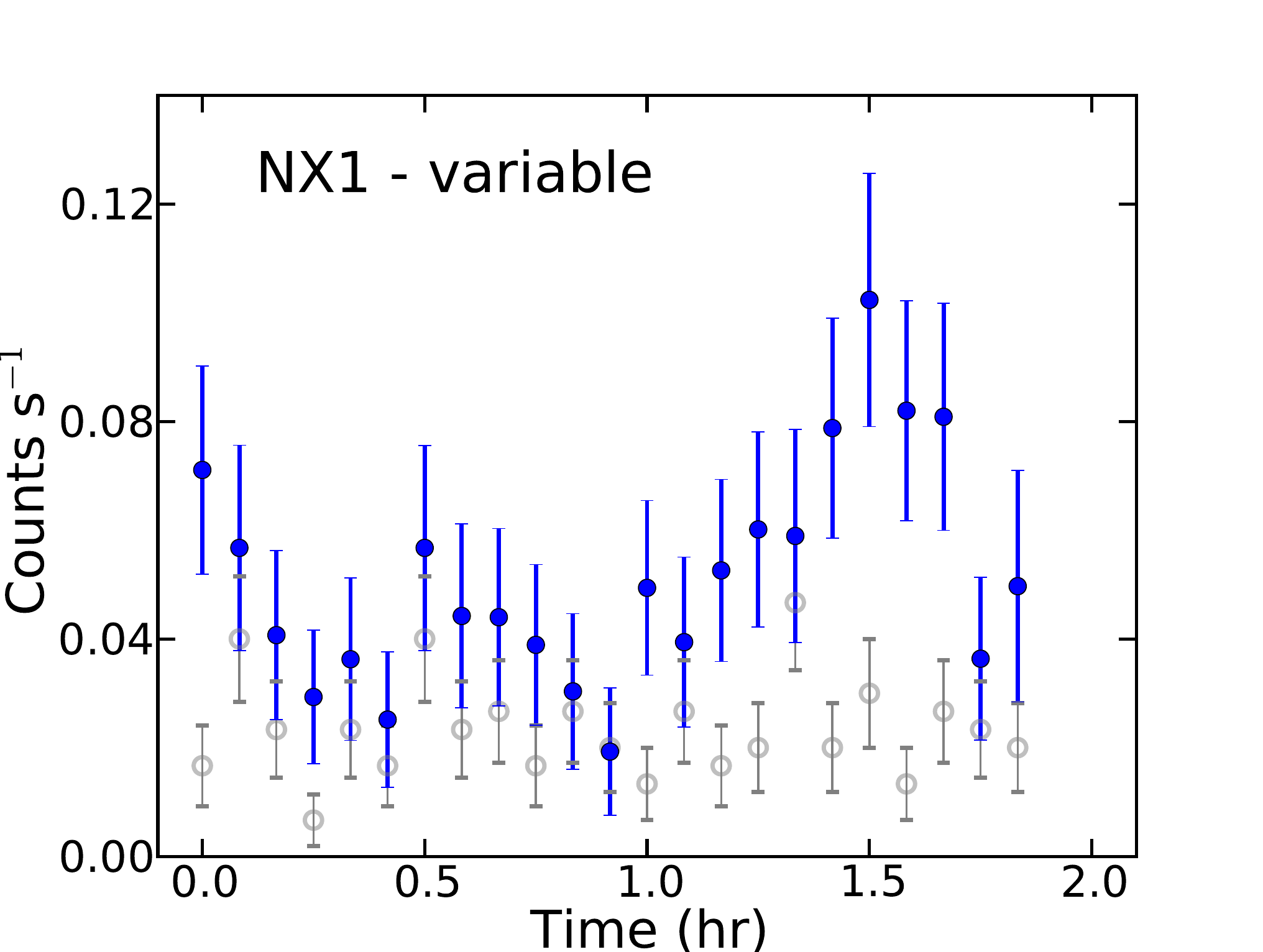}
\includegraphics[width=1.65in,viewport=28 25 530 410,clip]{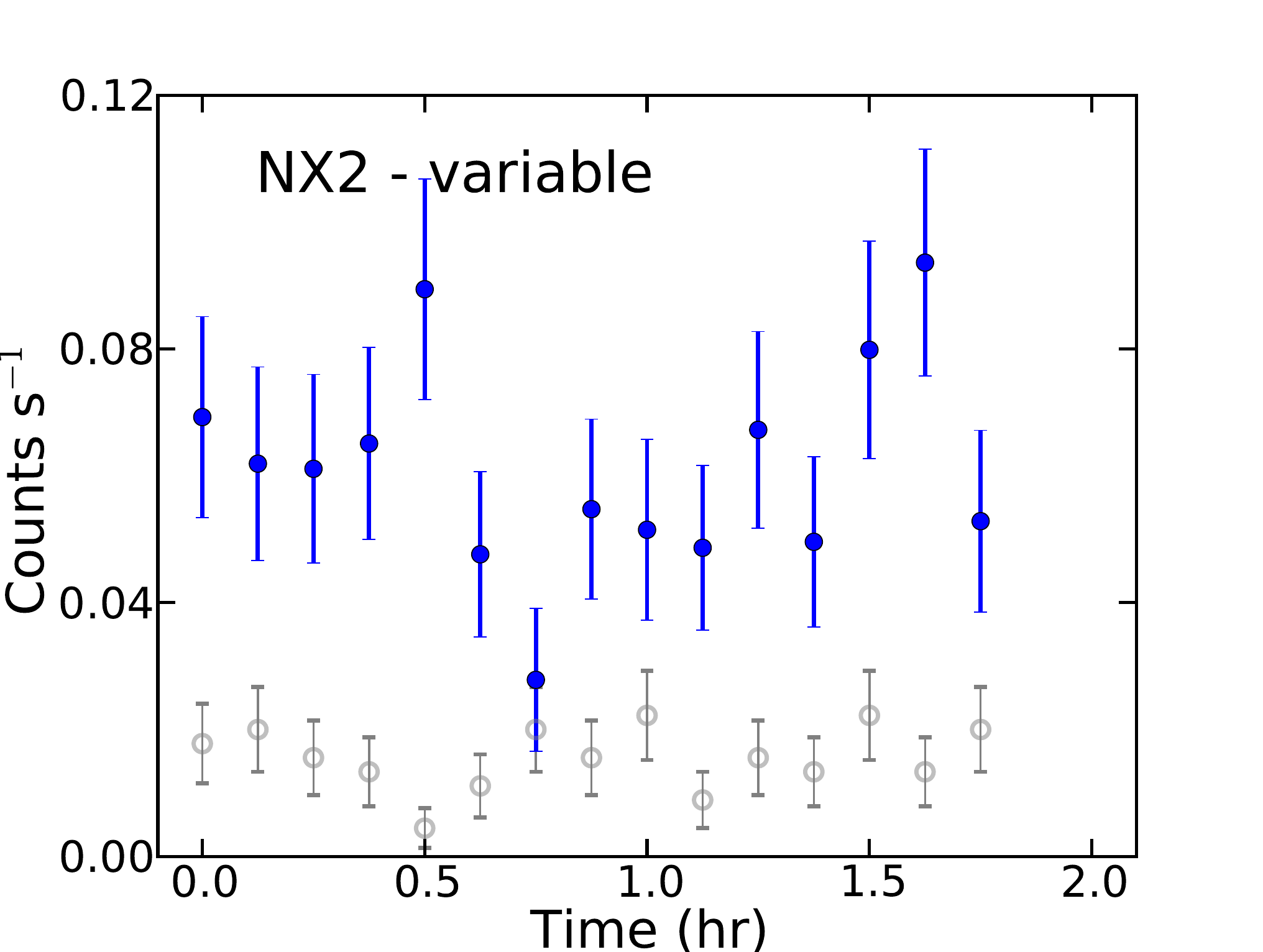}
\includegraphics[width=1.65in,viewport=28 25 530 410,clip]{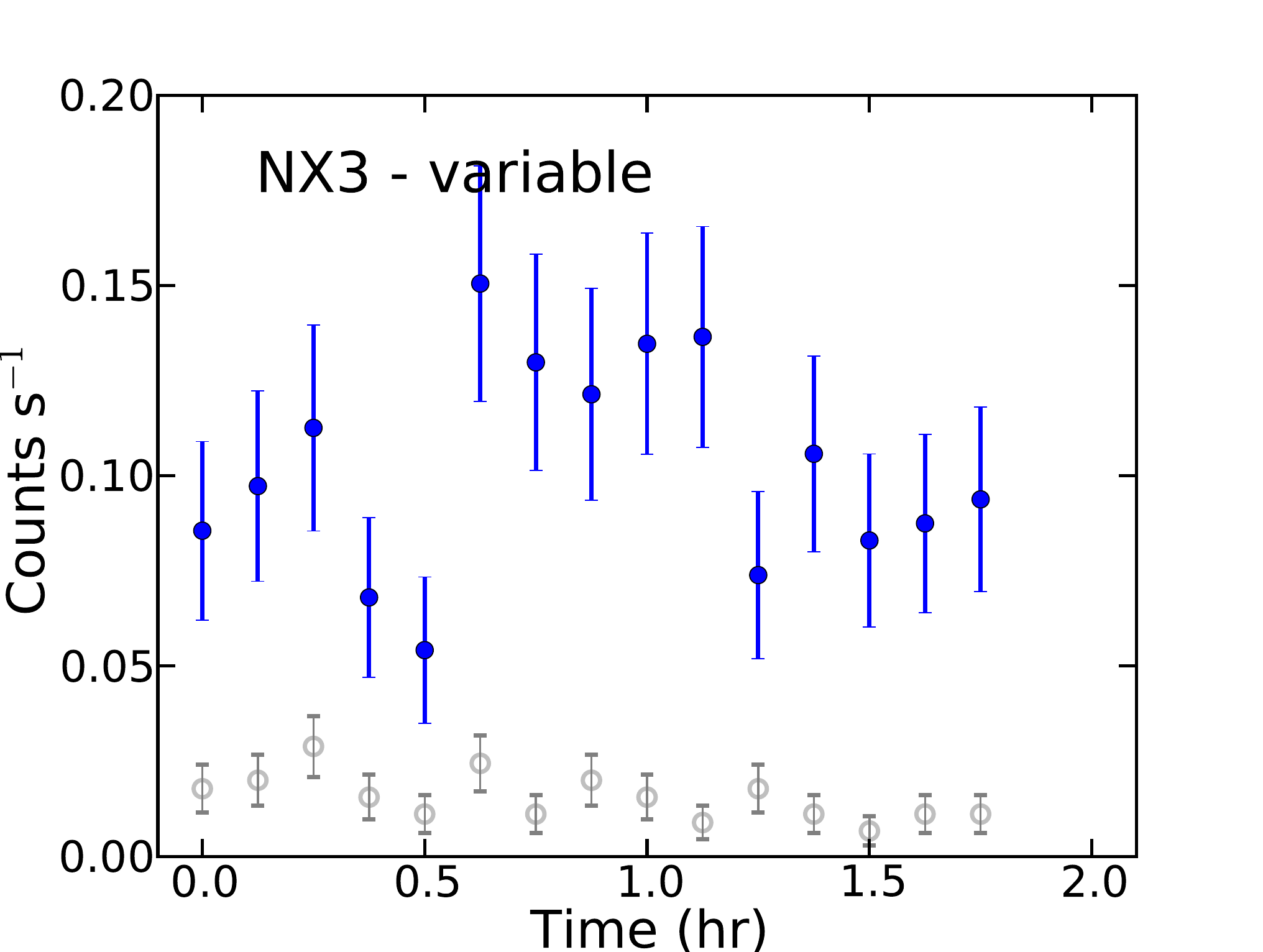}
\includegraphics[width=1.65in,viewport=28 25 530 410,clip]{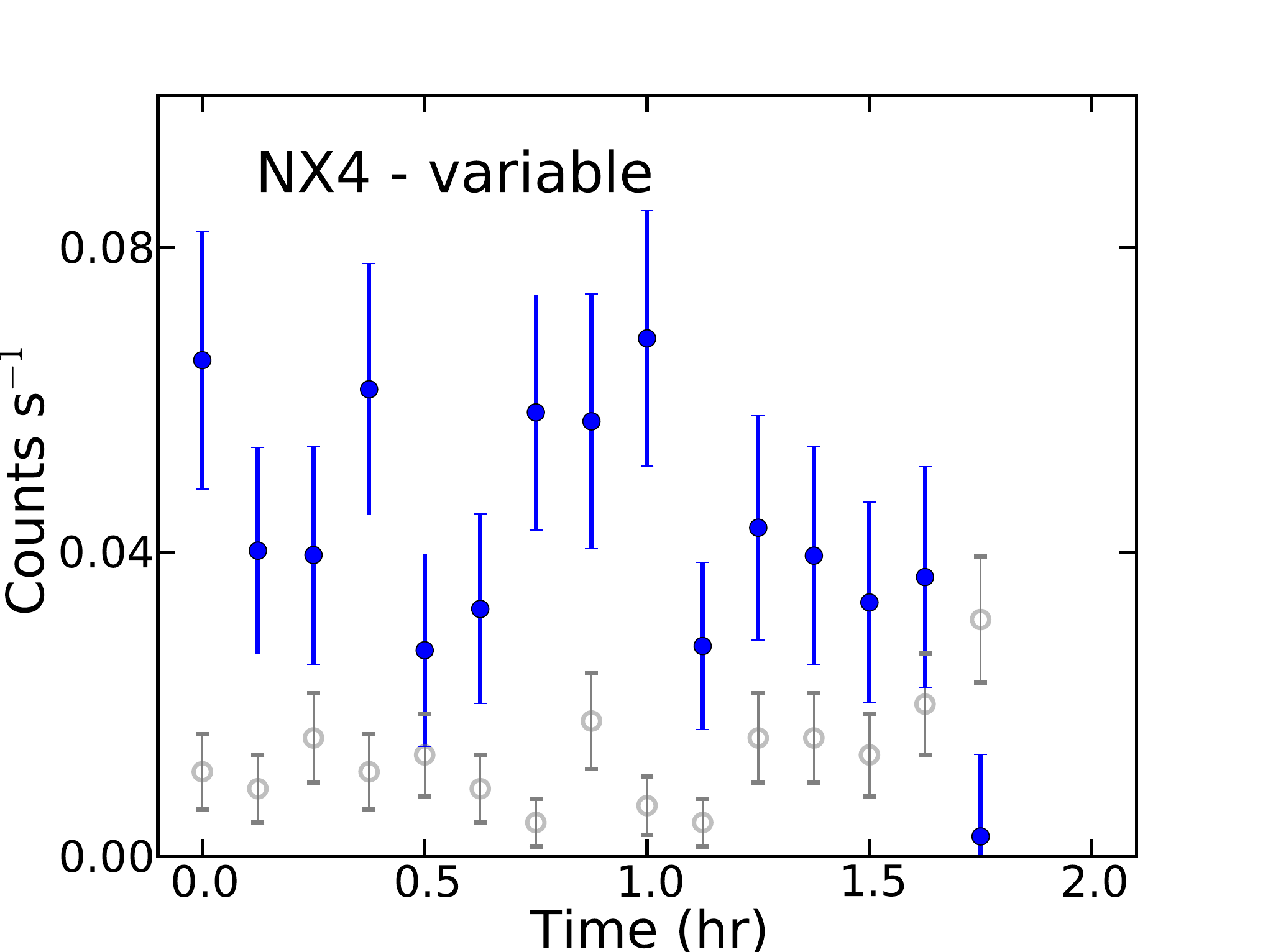}
\\
\includegraphics[width=1.76in,viewport=0 25 530 410,clip]{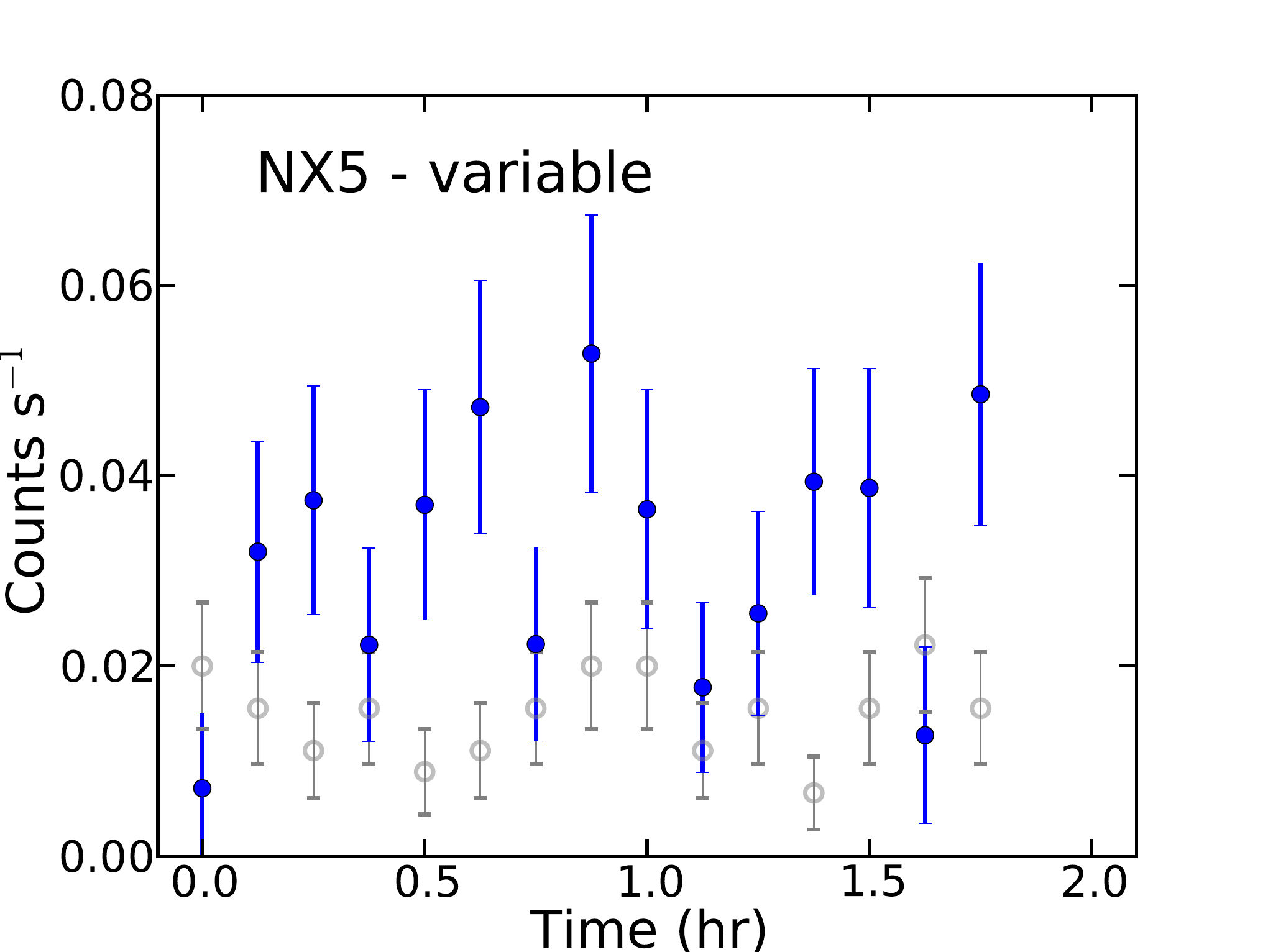}
\includegraphics[width=1.65in,viewport=28 25 530 410,clip]{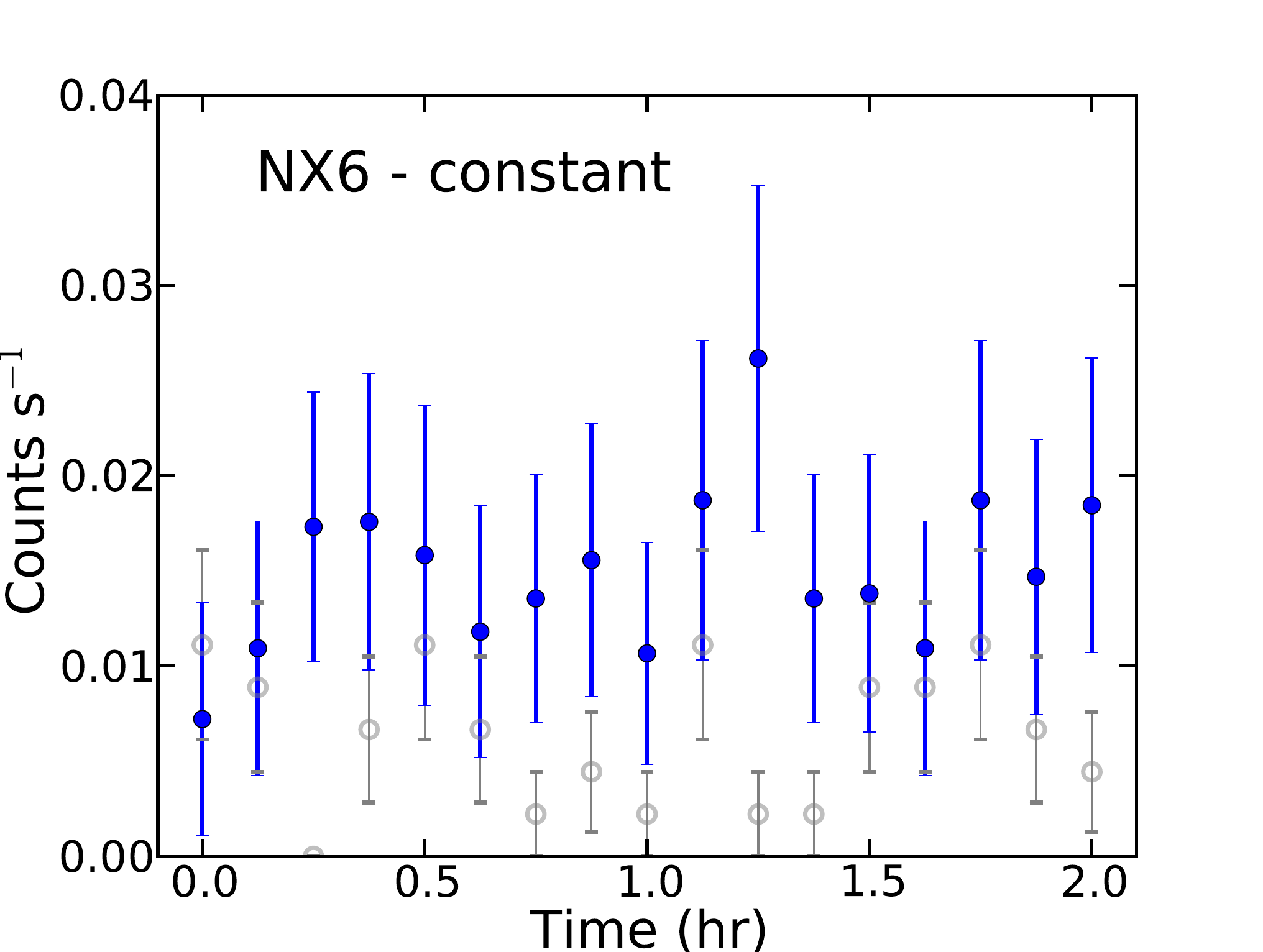}
\includegraphics[width=1.65in,viewport=28 25 530 410,clip]{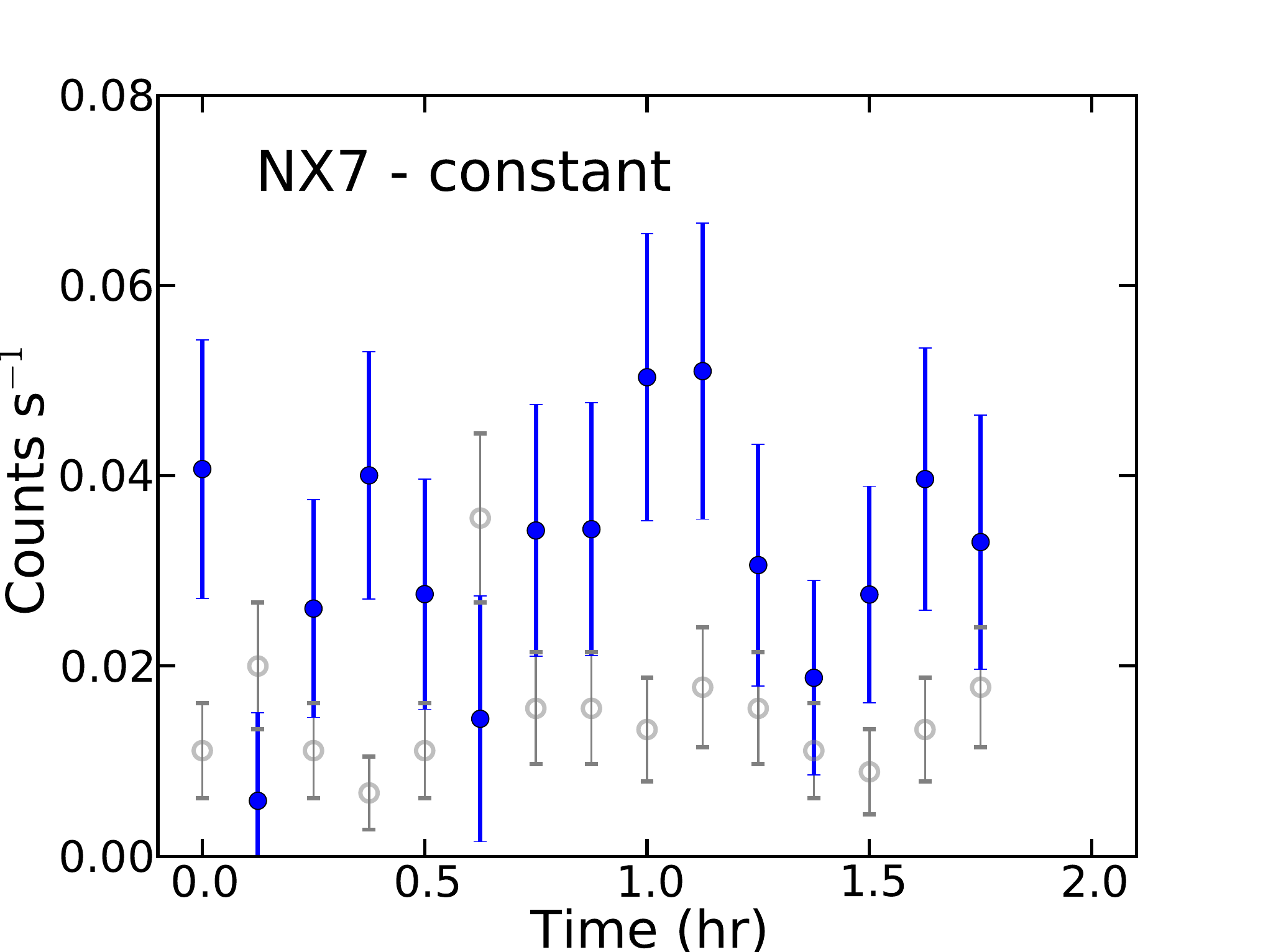}
\includegraphics[width=1.65in,viewport=28 25 530 410,clip]{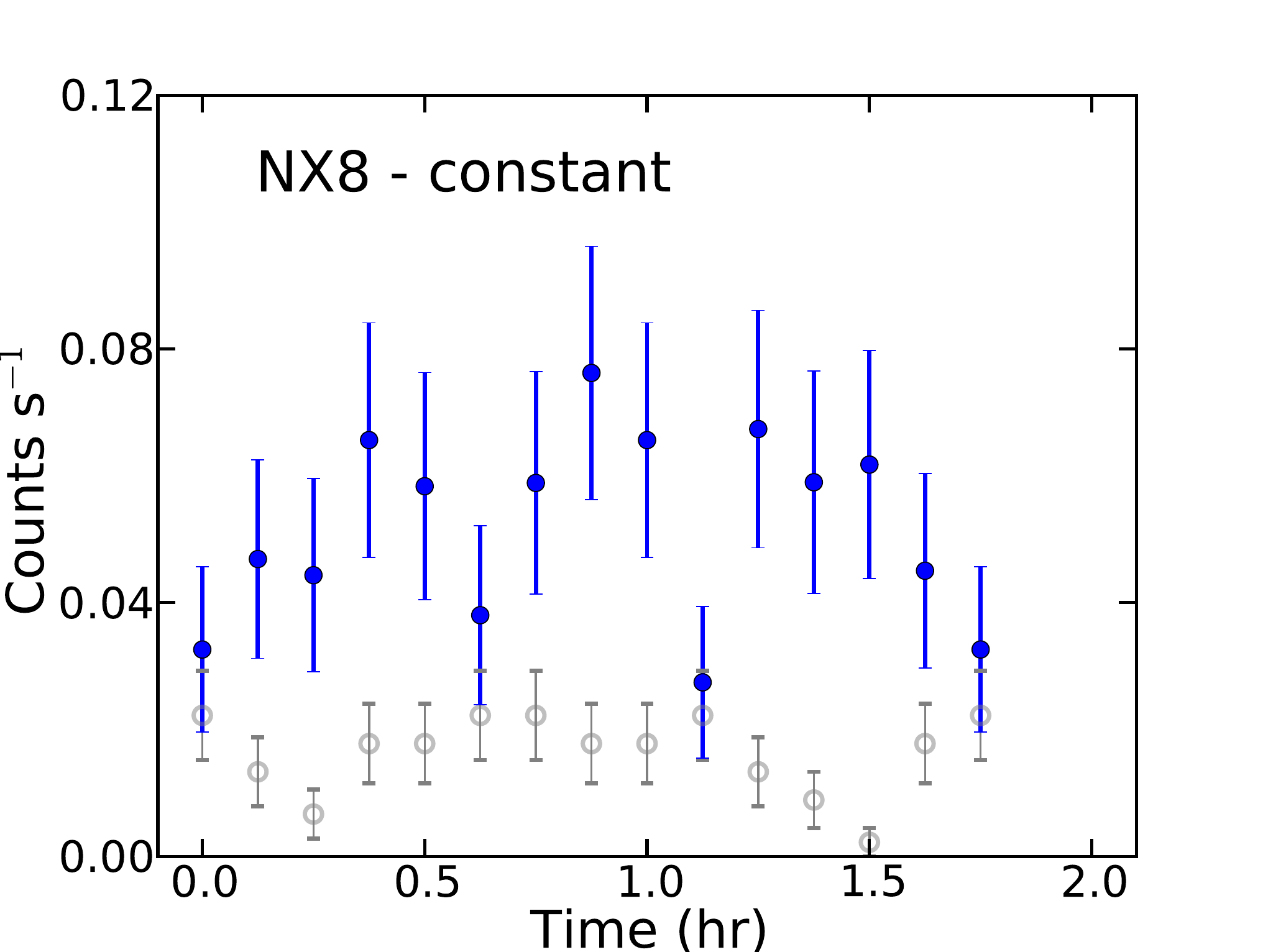}
\caption{The light curves for sources having 100 or more counts in PN (MOS1 for NX6). The background-subtracted light curves are in blue and the background-only light curves are in grey.
Countrate (counts per second) is on y-axis and time in hours is along the x-axis. The time bin size is set to 300 s for NX1 and 450 s for the rest in order to get 
$>$30 counts in each bin. Whether a source is variable or not according to the $\chi^2$ test is denoted on the upper left side. See \S\ref{sec:analyses.variability} for details.}
\label{fig:lc}
\end{figure*}

The X-ray light curves were extracted from the same regions as for the spectra from the filtered PN (MOS1 for NX6) 
event list in the 0.2--10 keV band with tasks {\tt evselect} and {\tt epiclccorr}.
The time bin size was set to 300 s (NX1) or 450 s (NX2--8) to get adequate signal-to-noise ratio in each bin.
For PN we selected only those events with $PATTERN$ keyword less than or equal to 4.
The light curves for NX1--8  are shown in Figure~\ref{fig:lc}.
The background-subtracted light curves are in blue and the background-only light curves are in grey.

% ---------------------------------------------------------------
% FIGURE: LIGHT CURVES
% ---------------------------------------------------------------
% ---------------------------------------------------------------

% ---------------------------------------------------------------
% FIGURE: LIGHT CURVES
% ---------------------------------------------------------------
\begin{figure}
\includegraphics[width=3.4in]{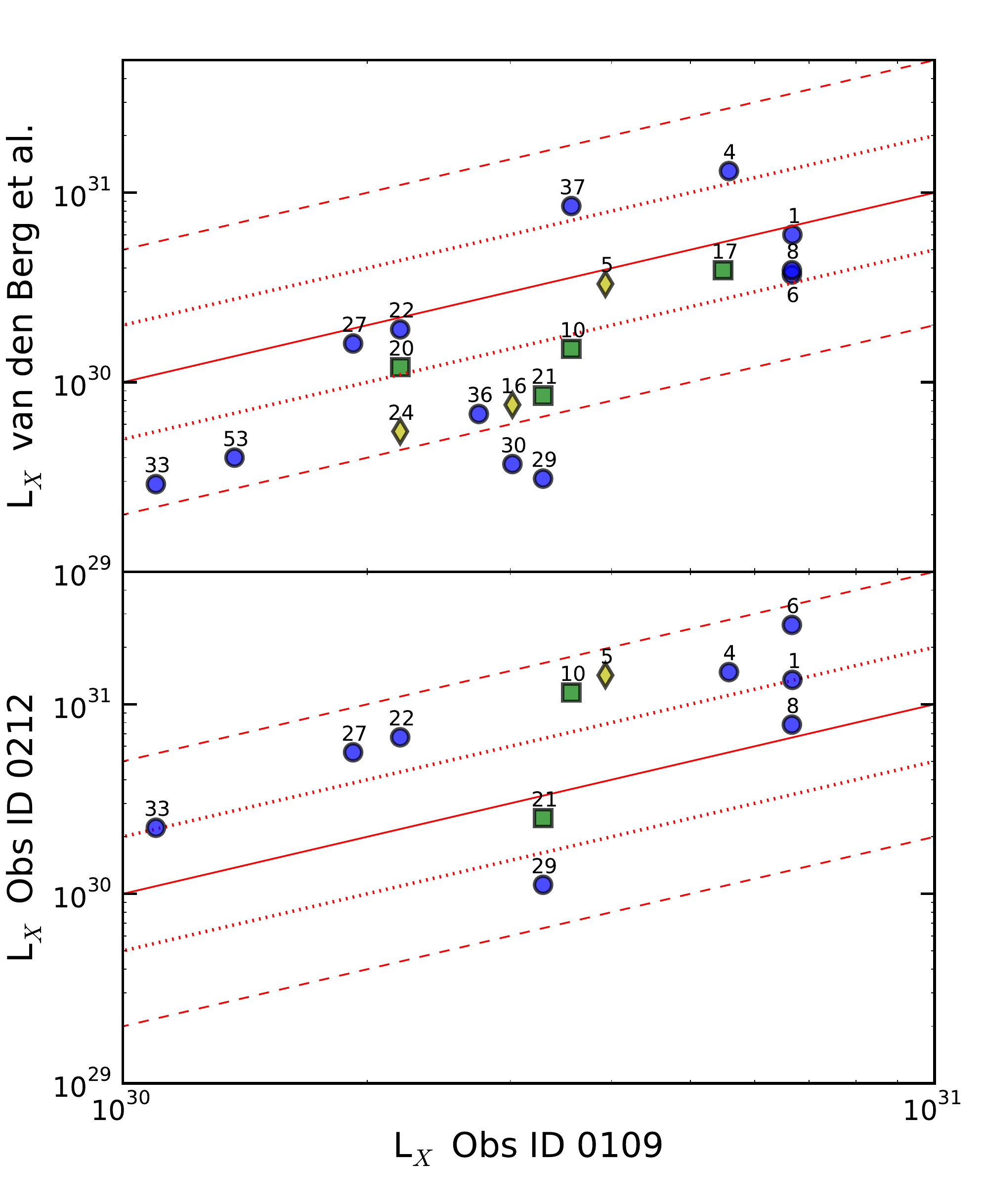}
\caption{Variability of X-ray members of M67 on 5-month (top) and 4-year (bottom) timescale shown using plots of the X-ray luminosities of sources 
in Obs ID 0109461001 versus X-ray luminosities in \citet{berg04} (top) and those in Obs ID 0212080601 (bottom) respectively. 
The solid red line marks equal luminosities in either quantity plotted, and the red dotted (dashed) lines denote variability of a factor of 2 (5). 
RS CVn or Algol systems are marked with blue circles, W UMa systems with green squares, and all other M67 members (all of which are yellow stragglers) are shown as yellow diamonds.
See \S\ref{sec:analyses.variability} for details. Note that the implied X-ray luminosities in the 0.2--7.0 keV energy band are used.}
\label{fig:lxvar}
\end{figure}
% ---------------------------------------------------------------

We used the $\chi^2$ test to check for the departure from white noise in the light curves, and selected sources having $\chi^2$ probability $<$25\% as significant variables.
% Note that this assumes Gaussian statistics for which grouping photon counts into bins with $\simeq$10 counts in each bin is necessary.
% Since the $\chi^2$ test does not give any measure of the timescale of variability, we calculated the autocorrelation function (ACF) 
% as a measure of timescale of variability. The ACF gives the magnitude of correlated structure in a time series \cite[e.g. ][]{feigelson2012}, and we used the ACF analysis tools implemented 
% in the statistical computing package, {\tt R}\footnote{http://www.r-project.org/}. We have adopted a slightly modified definition of the ACF (equation~\ref{eqn:acf}), 
% replacing photon counts with the photon countrate.
% 
% \begin{equation}
% ACF(k) = \frac{ \sum_{t=1}^{N-k} (R_t-\bar{R}) ~ (R_{t+k}-\bar{R}) }{ \sum_{t=1}^{N} (R_t-\bar{R})^2 }
% \label{eqn:acf}
% \end{equation}
% 
% \noindent where $\bar{R}$ is the mean photon countrate (Table~\ref{tab:xraysrc}, and $k$ is a positive integer called the lag time.
% Note that this estimator is biased since, in reality, $\bar{R}$ and the variance has been calculated from the data.
% Also, the ACF calculated here is to be considered only approximate since the variance on the countrate in each bin is not the same.
% 
Sources NX1 to NX5 are thus found to be variable, while the variability in NX6--8 is not significant.
% It might be the case that NX6--8 are variable on $\sim$minute timescales, but we do not have enough signal-to-noise 
% ratio to test such a hypothesis. 
Members of M67 showing a significant variability are NX1, NX4 and NX5, which also exhibit short-term autocorrelation.
For the RS CVn binary, NX1, the autocorrelation function shows a significant correlated structure for 
lag time k$\leqslant$2 implying variability timescale of $\sim$10 min. 

We also compared the X-ray luminosities of M67 members detected in the {\it XMM} observation ID 0109461001 with those in the {\it Chandra} 
observation from \cite{berg04} separated by five months, and between ID 0109461001 and ID 0212080601 to find the variability over two observations spaced by $\sim$4 years.
The comparison plots are shown in Figure~\ref{fig:lxvar}. 

The RS CVn-type systems, NX1, NX6, NX22, and XN27, have relatively stable X-ray emission between ID 0109461001 and \citeauthor{berg04}, but are variable by more than a factor of 
two between ID 0109461001 and ID 0212080601, while NX8 appears stable between all three observations.
NX4, NX16, NX33, NX36, NX37, and NX53 are variable by more than a factor of two between any of the two observations.
NX29 and NX30 are variable by more than a factor of five over the observations spaced by five months.
The W UMa-type system, NX21, varied by a factor of four between ID 0109461001 and Chandra observations.
NX10 is variable by more than a factor of two over the observations separated by months as well as years.
Both types of systems, RS CVn-type and W UMa-type, are known to be highly variable due to recurrent flaring activity.
The snapshot observations, which are not very long, could have caught them in different phases of their activity.
Three yellow stragglers, which are present in the multi-epoch data, also display average change in luminosity by a factor $>2$.
Figure~\ref{fig:lxvar} suggests that, the variability amplitudes of RS CVn-type sources, W UMa-type systems, and yellow stragglers are similar.

%--------------------------------------------------------------------------------------------
% NOTES ON SOURCE CLASSES
%--------------------------------------------------------------------------------------------
\section{NOTES ON INDIVIDUAL CLASSES OF STELLAR X-RAY SOURCES AND THEIR X-RAY LUMINOSITY FUNCTIONS} \label{sec:source_notes}
In Table~\ref{tab:classes}, we provide notes on the individual classes of stellar X-ray sources in M67.
We have used the orbital parameters for X-ray members from \cite{berg04}, who have in turn compiled them 
from \cite{mathieu1990}, \cite{latham1992}, \cite{berg2000}, and from unpublished work by D. Latham, R. Mathieu et al. 
For the names of the sources and their parameters, please see Table~\ref{tab:members}.

% ---------------------------------------------------------------
% TABLE: Notes on Individual Classes Of Stellar X-Ray Sources
% ---------------------------------------------------------------
\begin{table*}
\caption{Notes on Individual Classes Of Stellar X-Ray Sources}
\label{tab:classes}
\begin{tabular}{p{2cm}p{0.01cm}p{1in}p{4.8in}} % 3 cols
\hline \hline
Class / Type& & Source(s)                                & Notes\\
\hline
RS CVn$^a$  & 1) & NX6, NX8, NX36                           & Lie along subgiant branch. $P_{\rm orb}\lesssim10$ d, $e\lesssim0.1$.\\
            & 2) & NX1, NX22, NC29, RX26, RX46, CX78, CX88  & Proposed as RS CVn systems by \cite{berg04} and \cite{belloni98} based on the presence of Ca H \& K emission lines, X-ray spectrum, 
                                                         orbital period and circularisation. Lie along subgiant branch. $P_{\rm orb}<8$ d, $e\simeq0$.\\
            & 3) & NX27, NX64, RX28                         & Candidate RS CVn-type systems or their main-sequence analogs. Complete orbital solution not available. \\
\hline
Algol       & 1) & NX37                                     & Previously known Algol system. $P_{\rm orb}=1.1$ d.\\
            & 2) & CX157                                    & Candidate Algol-type system. Consists of an eclipsing system containing an F0-type and an early M-type main sequence star \citep{gokay2013}.\\
\hline
Contact binaries&   & NX10, NX17, NX20, NX21, NX53=CX61     & All are W UMa-type systems. $P_{\rm orb}\lesssim0.5$ d. 
                                                         NX10 and NX17 are eclipsing binaries.
                                                         NX61 had been assigned unknown membership probability by \cite{berg04}, and we confirm its M67 membership in this work.
                                                         NX10, NX17, NX21, and NX53 are on the main sequence. NX20 is close to the main-sequence turnoff.\\
\hline
CV          & 1) & RX35=NX73                                & \cite{belloni98} use $B-V=0.38$ from \cite{sanders77} and hardness of the X-ray spectrum to claim that this could be an accreting white dwarf system. 
                                                         Orbital parameters are unknown. However, EIS has $B-V=0.79$, and our hardness ratios are $HR1=0.66\pm0.20$, $HR2=-0.48\pm0.30$. 
                                                         The position in the HR diagram and absence of soft and hard X-ray components therefore argue against the CV hypothesis for this source.\\
            & 2) & NX42                                     & AM Her-type CV, EU Cnc. 
                                                         {\it ROSAT} detection is only below 0.4 keV \citep{belloni98}, but \cite{berg04} find a relatively high hardness ratio.
                                                         In this work we have $HR1=-0.50\pm0.33$ and $HR2=0.23\pm0.46$, consistent with the high hardness ratios typical of magnetic CVs. \\
%                                                          For a thermal brehmsstrahlung spectrum (0.1 keV), $L_X=2.1\times$10$^{29}$ erg s$^{-1}$, which is a factor of two lower than 
%                                                          the value reported by \citeauthor{berg04}, suggesting variable emission on a 5-month timescale.\\
\hline
Blue Stragglers  & 1) & NX37                          & 1.1 d orbit for the inner binary of a possible triple system \citep[][]{goranskij1992,berg2001}.
                                                         \cite{berg2000} find the spectral type to be F5IV, and \cite{berg2001} suggest that the X-ray emission could be due to magnetic activity in the rapidly rotating subgiant.\\
                 & 2) & CX94$^b$                      & Noted as a blue straggler in \cite{sandquist2003}, but with the EIS $B-V$ color, it is very close to main-sequence turnoff.
                                                         \cite{berg04} noted that CX94 shows no radial-velocity variations over six observations spanning 3923 days.\\
                 & 3) & CX95$^b$                      & $P_{\rm orb}=4913$ d and $e=0.3$. % BS 111 in Liu et al. 2008
                                                         The primary has spectral type F6V and a projected rotation speed, v\,{\it sin}($i$)$\simeq$20 km s$^{-1}$ \citep{latham1996,liu2008}.
                                                         \citeauthor{berg04} postulate that CX95 could have an undetected close binary, hence explaining the X-ray emission.\\
\hline
Yellow Stragglers& 1) & NX5                          & Binary system with a G4 giant and a cool white dwarf, with a circular orbit having $P_{\rm orb}=43 d$.
                                                         Ca II H,K emission lines are present \cite{pasquini1998}. 
                                                         Circularized orbit suggests strong tidal forces between the two components of the system, but the cool temperature of the white dwarf precludes mass transfer.\\
                 & 2) & NX16, NX24                   & $P_{\rm orb}=1495$ d and 698 d, and $e=0.32$ and 0.11 respectively. 
                                                       Tidal interaction therefore not strong enough to explain their X-ray emission, $L_X>10^{30}$.\\
\hline
Peculiar System  &    & NX4                                 & Lies about one mag below the sub giant and giant branches in the $V/B-V$ diagram (Figure~\ref{fig:hr}).
                                                         The X-ray spectrum and variability suggest that the X-ray emission is coronal.\\
\hline
New M67 Member   &    & NX75                                & Associated with the V$=$15 mag optical counterpart, E4630. 
                                                         Orbital parameters and hardness ratios are not well constrained. 
                                                         We estimate the spectral type of this source as K4V and a distance of 860 pc based on archival multiwavelength photometry, 
                                                         and thus suggest that it may be an RS CVn type.
                                                         Further optical spectroscopy and radial velocity measurements are needed for understanding its chromospheric activity and 
                                                         for deriving its orbital parameters.\\
\hline
Other Members with unknown class & 1) & NX45, CX67, CX94 & Lie along the F/G/K main sequence and are known binaries with undetermined orbital solutions.
                                                                 NX45 has a spectral type G0V and is close to the main-sequence turnoff.\\
                                 & 2) & NX62, CX62, CX73, CX77, CX80, CX82 & Lie along the M dwarf part of the main sequence of M67.
                                                                        These members are possibly spun-up binaries with active chromospheres where both members are main sequence stars.\\
\hline
\multicolumn{4}{p{7in}}{(a) Sources with $P_{\rm orb}\lesssim10$ d are expected to have near-circularized orbits \citep[e.g. ][]{latham2007}. 
(b) CX94 and CX95 are within the fov of the {\it XMM-Newton} observations, but below the detection threshold.}
\end{tabular}
\end{table*}
% ---------------------------------------------------------------

\begin{figure}
\includegraphics[width=3.6in]{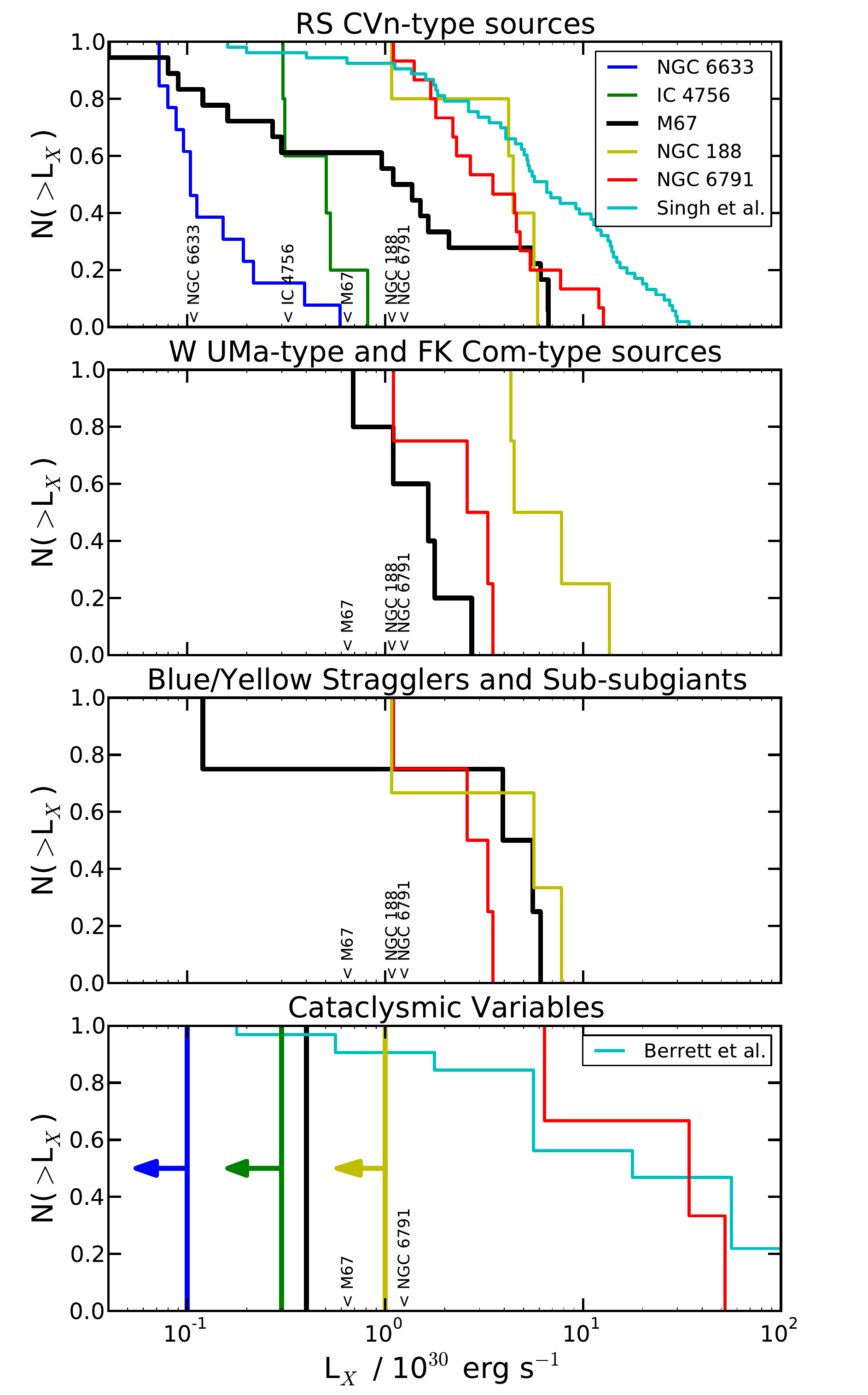}
\caption{The X-ray luminosity functions for RS CVn-type sources (top panel), contact and coalesced binaries (second panel from top), stragglers and other sources having 
anomalous locations in the color-magnitude diagram (third panel from top), and CVs (bottom panel) in different star clusters: 
NGC 6633, IC 4756 (0.7 Gyr), M67 (4 Gyr), NGC 188 (7 Gyr), and NGC 6791 (8 Gyr).
The sensitivity threshold for the different star clusters below which the source counts are incomplete are shown towards the bottom of each panel.
See \S\ref{sec:source_notes} for details.}
\label{fig:lum_function}
\end{figure}

In Figure~\ref{fig:lum_function} we plot the X-ray luminosity functions (XLFs)​ ​of various stellar types in M67 and compare compare them 
with the XLFs of other open star clusters of intermediate-to-old age.
The star clusters used for comparison with M67 (4 Gyr; Table~\ref{tab:members}) are: NGC 6633, IC 4756 \citep[0.7 Gyr; ][]{briggs2000}, NGC 6791 
\citep[8 Gyr; ][]{berg13}, and NGC 188 \citep[7 Gyr; ][]{belloni98,gondoin2005}.
The clusters in general have different median values for the X-ray luminosity for different stellar types.
The XLF of the RSCVns in M67 is shown in the top panel of Figure~\ref{fig:lum_function} along with that in the other ​clusters. 
The XLF of field RS CVn binaries from \cite{singh1996} is also shown for guidance.
% The clusters, in general have different median values for the X-ray luminosity.
In general, RS CVn-type systems are more numerous at the low luminosity end, and there exist very few systems above $L_X=10^{31}$ erg s$^{-1}$.
It is evident that the highest luminosity of RS CVn-type systems seen in the clusters is a function of the age of the clusters.
Older clusters appear to have larger number of brighter luminosity systems.
In other words, the XLFs indicate that the RS CVn-type systems get brighter as they age.
It is also evident from Figure~\ref{fig:lum_function} that the RS CVn XLFs of M67 and the \citeauthor{singh1996} sample have a much shallower slope than 
those of the other clusters, which may be due to two distinct distributions of RS CVn-type systems in open clusters\footnote{We could have also missed 
a population of low luminosity RS CVns due to the limit of our sensitivity, but this would only partly explain the shallow slope.}.
This is suggestive of the intuitive notion that, as the clusters age, more active binary systems and active stars are produced \citep[see also ][]{verbunt99,berg13a}. 
This picture is consistent with younger systems having lower luminosity and older ones being more luminous.
The shallower slope of and the presence of several $L_X>10^{31}$ erg s$^{-1}$ systems among the field population advocate that these are old systems which may have been 
kicked out of their parent clusters.

The second panel from the top of Figure~\ref{fig:lum_function} shows the XLFs of WU UMa-type and FK Com-type systems 
in M67, NGC 188, and NGC 6791.
Down to the sensitivity of X-ray surveys in NGC 6633 and IC 4756, these clusters do not have any definite detections of contact or coalesced binaries.
From Figure~\ref{fig:lum_function} we see that the slopes of the contact binary XLFs are steep, and there are several systems having X-ray luminosities greater than $10^{30}$ erg s$^{-1}$.
The WU UMa-type and FK Com-type systems in NGC 188 are more luminous ($L_X>4\times10^{30}$ erg s$^{-1}$) compared to those in M67 and NGC 6791.
This is surprising given that NGC 188 is neither the most massive (in fact, it is the least massive among the old clusters) nor the oldest cluster among the sample.

The third panel from the top of Figure~\ref{fig:lum_function} shows the XLFs of stragglers and peculiar systems.
The brightest ends of the XLFs of M67, NGC 188, and NGC 6791 are steep.
Majority of the straggler and peculiar systems known in these clusters span a narrow range in X-ray luminosity, $3\times$ to $6 \times 10^{30}$ erg s$^{-1}$.
This could indicate a fine-tuning of X-ray emission for the brightest systems, which needs further investigation.
At the low luminosity end, incompleteness due to survey sensitivities is apparent.
As such, among the old clusters, the XLFs appear to be independent of the cluster age.

The bottom panel of Figure~\ref{fig:lum_function} shows the XLFs of cataclysmic variables.
The XLF of a mixed sample of polars from a variety of binaries from \cite{barrett1999} is also shown for guidance.
NGC 6633, IC 4756, which are 0.7 Gyr old, do not have any CVs.
M67 (4 Gyr old) has only one CV, the polar in quiescence EU Cnc, which has 0.2--7 keV X-ray luminosity of $<10^{30}$ erg s$^{-1}$, while the older 
cluster NGC 6791 contains three CVs with $L_X\gg10^{30}$ erg s$^{-1}$.
NGC 188, although an old cluster of age 7 Gyr, does not have any CVs brighter than 10$^{30}$ erg s$^{-1}$.
Given this information, the production of bright CVs appears to be correlated with the mass of the cluster.
NGC 6791 is about 10 times more massive than M67, and M67 is about twice as massive as NGC 188 \citep[][and references therein]{pandey1987,berg13}.
\cite{berg13} find that the number of CVs per unit mass is consistent with the field, and conclude that the three CVs which are confirmed members of NGC 6791 are primordial.
Unlike RS CVn systems, in the case of CVs we cannot ascertain the evolution of the XLF with age.
However, the mass of the parent cluster has to play an important role.
More CVs are produced as clusters age, and it is plausible that these systems also get brighter with age.
In this picture, the \cite{barrett1999} sample, which contains numerous polars at high luminosity ($L_X>10^{31}$ erg s$^{-1}$), could be very old systems.
The shallow slope of the \citeauthor{barrett1999} XLF is likely due to the bimodal distribution of polars in X-ray luminosity as noted by \citeauthor{barrett1999}, which may be related to age of the systems.

% 
% XLFs
% compare the properties
% lesser or more content
% are they at the low lum end or high end? e.g. RS CVn are at low lum end

% ---------------------------------------------------------------
% FIGURE: LUMINOSITY FUNCTIONS
% ---------------------------------------------------------------
% ---------------------------------------------------------------

%--------------------------------------------------------------------------------------------
% SUMMARY
%--------------------------------------------------------------------------------------------
\section{SUMMARY} \label{sec:summary}

Our analysis of two {\it XMM-Newton} observations of the old open cluster, M67 has led to the detection of 25 members of M67, of which 24 
have been detected in previous observations with {\it Chandra} \citep{berg04} and {\it ROSAT} \citep{belloni93,belloni98}. 
One X-ray source is a newly-detected member. 
Based on present and older observations we have compiled an updated list of 43 X-ray members of M67 (Table~\ref{tab:members}), 16 of
which are likely to be RS CVn or related binaries with circularised or near-circularised orbits having periods P$_{orb}$ $\lesssim$10 d. 
We also detected five contact binaries with P$_{orb}\lesssim$0.5 d.
% The cataclysmic variable EU Cnc is also detected, but although it was previously considered to be a member of M67, has a very low membership probability (10\%) based on proper motion \citep{yadav08} 
% and a non-member based on stellar apexes method \citep{vereshchagin14}. 
X-ray emission from three yellow stragglers and two blue stragglers detected here is not well understood.
Another peculiar source, HU Cnc, lies below the intersection of the sub-giant and red giant branches in the HR diagram (Figure~\ref{fig:hr}), 
and may be a coronal emitter.
The only cataclysmic variable known in M67, the AM Her-type system EU Cnc, is detected in the {\it XMM-Newton} observations considered in this work.
Fourteen X-ray members do not have any definitive classification; some of these are known binaries with undetermined orbital solutions.
Spectroscopic follow up observations and radial velocity monitoring in the optical and at other wavelengths is required for elucidating the nature of these sources.
M67 lies within the field of {\it Kepler}'s K2 observing campaign\footnote{http://keplerscience.arc.nasa.gov/K2/}, and the optical light curves from this campaign will be useful for calculating orbital parameters of at least some of these systems.
We already have an existing program to study the ultraviolet emission from M67 members and to compare it with their X-ray emission 
(Subramaniam et al., in prep). 
We found X-ray variability among M67 members on a five-month timescale and a $\sim$4-year timescale 
by comparing the {\it XMM-Newton} observations and Chandra observations from \cite{berg04}.
The fractional variability of RS CVn-type sources, W UMa-type systems, and yellow stragglers are found to be quite similar, factor of $\sim$3 on months and years timescales.
Finally, we have studied the X-ray luminosity function of RS CVn-type (and other types) members in M67. 
A comparison with that of other intermediate-to-old open clusters (Figure~\ref{fig:lum_function}) shows an increased number of higher luminosity stars 
in the older clusters, thus suggesting that more active binary systems and active stars are likely produced with the aging of clusters. 
% The data for the other types is insufficient to draw any firm conclusions.
Deeper and more sensitive X-ray and optical observations of intermediate-to-old star clusters are encouraged in order to find fainter sources 
and thus extend the luminosity functions of the different classes of X-ray sources to the faint end.

%--------------------------------------------------------------------------------------------
% ACKNOWLEDGEMENTS
%--------------------------------------------------------------------------------------------
\section*{ACKNOWLEDGMENTS}
KM would like to thank Himali Bhatt and Dom Walton for useful discussions and guidance on X-ray data processing.
We thank the anonymous referee for the valuable suggestions which greatly improved the manuscript.
This research has made extensive use of Vizier, SIMBAD, and SDSS.

%--------------------------------------------------------------------------------------------
% BIBLIOGRAPHY
%--------------------------------------------------------------------------------------------

\end{document}